\documentclass[aps,twocolumn,showpacs,preprintnumbers,floats]{revtex4}\def\@cite#1#2{\textsuperscript{[{#1\if@tempswa , #2\fi}]}}

\usepackage{graphicx}
\usepackage{amsmath}
\usepackage{bm}
\usepackage{amsfonts}
\usepackage{amssymb}
\usepackage{color}
\usepackage{subfigure}
\usepackage{epsfig}
\usepackage{morefloats}
\usepackage{multirow}
\usepackage{graphicx,booktabs}
\usepackage{mathrsfs}
\usepackage{txfonts}
\usepackage{pifont}
\usepackage{indentfirst}
\usepackage{graphicx,booktabs}
\usepackage{longtable,lscape}
\usepackage[figuresright]{rotating}
\usepackage[colorlinks, citecolor=blue,anchorcolor=red,menucolor=red, linkcolor=red,filecolor=red,urlcolor=blue,frenchlinks=red]{hyperref}

\begin{document}

\title{Charmonia in an unquenched quark model}

\author{Qian Deng$^{1}$, Ru-Hui Ni$^{1}$~\footnote{Ru-Hui Ni and Qian Deng contributed equally to this work.} \footnote{E-mail: ruhuini@foxmail.com}, Qi Li$^{2}$, and
Xian-Hui Zhong$^{1,3}$~\footnote {E-mail: zhongxh@hunnu.edu.cn}}

\affiliation{ 1) Department of Physics, Hunan Normal University, and Key Laboratory of Low-Dimensional Quantum Structures and Quantum Control of Ministry of Education, Changsha 410081, China }

\affiliation{ 2)  School of Science, Tianjin Chengjian University, Tianjin 300000, China}

\affiliation{ 3)  Synergetic Innovation Center for Quantum Effects and Applications (SICQEA),
Hunan Normal University, Changsha 410081, China}


\begin{abstract}
In this work, we study the charmonium spectrum within an unquenched quark model
including coupled-channel effects. In couple-channel calculations, we include all of the
opened charmed meson channels with the once-subtracted method, meanwhile adopt a suppressed factor to
soften the hard vertices given by the $^3P_0$ model in the high momentum region.
We obtain a good description of both the masses and widths for the well-established
states in the charmonium spectrum.
Furthermore, we give predictions for the higher
$S$-, $P$- and $D$-wave charmonium states up to mass region of
$\sim 5.0$ GeV. The magnitude of mass shifts due to the coupled-channel
effects is estimated to be about $10s$ MeV. Although many decay channels are opened for the
higher charmonium states, they are relatively narrow states.
Their widths scatter in the range of $\sim 10s-100$ MeV.
Many charmonium-like states, such as $\chi_{c1}(3872)$, $\chi_{c1}(4274)$,
$\chi_{c0}(3915)$, $\chi_{c0}(4500)$, $\chi_{c0}(4700)$, $X(4160)$, $X(4350)$,
$Y(4500)$, and $\psi(4660)$/$Y(4710)$, can be accommodated by
the charmonium spectrum when the unquenched coupled-channel effects are carefully considered.

\end{abstract}


\maketitle

\section{Introduction}

Since the $X(3872)$ was first observed at Belle in 2003~\cite{Belle:2003nnu}, many charmonium-like states have been observed in the past two decades at experiments such as \emph{BABAR}, Belle, LHCb, and BESIII, etc..
Sixteen states $\chi_{c1}(3872)$ (known as $X(3872)$), $\chi_{c0}(3860)$, $\chi_{c0}(3915)$, $X(3940)$, $X(3960)$, $\chi_{c1}(4140)$, $X(4160)$,
$\psi(4230)$, $\chi_{c1}(4274)$, $X(4350)$, $\psi(4360)$, $\chi_{c0}(4500)$,
$X(4630)$, $\psi(4660)$, $\chi_{c1}(4685)$ and $\chi_{c0}(4700)$, are listed
in the Review of Particle Physics 2023 (RPP 2023) by the
Particle Data Group (PDG)~\cite{ParticleDataGroup:2022pth}.
Recently, the center-of-mass energies of BESIII experiments have been extended to 4.95 GeV, which bring new opportunities for searching
for the higher charmonium states. Lately, in the
higher mass region three new vector states $Y(4500)$~\cite{BESIII:2022joj,BESIII:2023cmv},
$Y(4710)$~\cite{BESIII:2022kcv} and $Y(4790)$~\cite{BESIII:2023wsc} were observed at BESIII.
Since these charmonium-like states contain a $c\bar{c}$ pair, they may be candidates
of charmonium states. However, as a $c\bar{c}$ assignment
the observed properties, such as mass, for some of them are out of the conventional
quark model expectations. They may be also candidates of exotic states,
such as multi-quark state, hadronic molecule, or hybrid.
About the nature of the charmonium-like states, there are many debates in the literature.
To know about the experimental and theoretical status of the charmonium-like states, one can refer to the review works~\cite{Dong:2021bvy,Ali:2017jda,Esposito:2016noz,Olsen:2017bmm,Lebed:2016hpi,Liu:2019zoy,
Chen:2016qju,Brambilla:2019esw,Chen:2022asf,Guo:2017jvc}.

As we know that the exotic states, such as tetraquark states, with normal quantum
numbers can hardly be distinguished from the normal ones, in order to search
for exotic states, one need have a good knowledge of the normal hadron spectrum.
In the charmonium spectrum, the low-lying states can be very well described
by various quark models with linear confinement potentials, such as the famous nonrelativistic
Cornell model~\cite{Eichten:1974af,Eichten:1978tg}, relativized Godfrey-Isgur model~\cite{Godfrey:1985xj,Barnes:2005pb},
relativistic quark model~\cite{Ebert:2011jc}, and so on. However,
for the description of higher charmonium states, the conventional quark model
will be questionable. The predictions often deviate from the observations.
These discrepancies may be caused by the quantum fluctuation, i.e., the creation and annihilation of
$q\bar{q}$ pairs in vacuum, which will bring an important effect.
That is the so-called ``coupled-channel effect"  through virtual
charm meson loops. This effect would become essential for some higher excited states.

The unquenched coupled-channel effects for charmonium states were evaluated within various
coupled-channel models based on different strategies in the literature, such as Refs.~\cite{Ferretti:2021xjl,Ferretti:2020civ,Duan:2021alw,Chen:2023wfp,Heikkila:1983wd,
Pennington:2007xr,Eichten:1978tg,Ortega:2019tby,Fu:2018yxq,Ferretti:2018tco,Li:2009ad,Barnes:2007xu,
Kalashnikova:2005ui,Anwar:2021dmg,Kanwal:2022ani,Zhou:2013ada,Ferretti:2013faa,Bruschini:2021cty,Lu:2017yhl}.
It is found that the masses of the higher excited states are often lowered by the coupled-channel effects.
Thus, some screened potential models, which roughly includes such unquenched effects,
were adopted to deal with the mass spectrum in the literature~\cite{Dong:1996ci,Ding:1993uy,Li:2009zu,
Li:2009ad,Deng:2016stx,Wang:2019mhs}.
The screened potential model results show that the mass suppression tends to be
strengthened from lower levels to higher ones, compared to the quenched potential model.
Both the screened potential model and coupled-channel model
seem to have the similar global features in describing the charmonium spectrum~\cite{Li:2009ad}.
In theory, to carry out unquenched calculations of charmonium spectrum, the best way is Lattice QCD.
There are some attempts in the literature~\cite{Wilson:2023anv,Wilson:2023hzu,Dudek:2007wv,HadronSpectrum:2012gic,DeTar:2011nn,Bali:2011rd,Lang:2015sba}.

For the aspect of phenomenology, one should face several challenges~\cite{Ni:HL} to seriously carry out a study of the charmonium spectrum
within the unquenched framework.
i) The first challenge is how to select the coupled channels. There are many coupled channels for
a charmonium state. However, only the low-lying channels
$D^{(*)}D^{(*)}$ and $D_s^{(*)}D_s^{(*)}$ are included in most of the calculations.
Can the contributions of the other higher channels be neglected for the high charmonium states?
ii) The second challenge is how to evaluate the coupled channel effects in the high momentum region.
Where the nonphysical contributions
may be involved due to the hard vertices given by the phenomenological models,
such as the $^3P_0$ model~\cite{Micu:1968mk,LeYaouanc:1972vsx,LeYaouanc:1973ldf}. There may be too large mass correction of
the coupled channel effects with a bare vertex.
iii) The third challenge is how to obtain a unified description of both
the masses and widths for the whole charmonium spectrum within the coupled-channel framework.
Usually, most of the works only care about mass corrections due to the coupled-channel
effects, a comprehensive description of the strong decays together with the mass
spectrum is very scarce.

In this work, we study both the masses and widths for the high charmonium spectrum
within a unified framework, where
the unquenched coupled-channel effects are included by combining
the semirelativistic linear potential model with the
$^3P_0$ model~\cite{Micu:1968mk,LeYaouanc:1972vsx,LeYaouanc:1973ldf}.
Two strategies are adopted to overcome the challenges existing in the
unquenched framework as mentioned above.
(i) First, in the coupled-channel calculations, we include all possible Okubo-Zweig-Iizuka (OZI) allowed charmed meson channels with mass thresholds below the bare $c\bar{c}$ states together with the nearby virtual channels. The contributions from the other virtual channels (whose mass thresholds are significantly above the bare $c\bar{c}$ states) are subtracted from the dispersion relation by redefining the bare mass with the once-subtracted method suggested in Ref.~\cite{Pennington:2007xr}. In other words, the mass shift of a bare state is entirely given by the opened and nearby virtual channels. (ii) Second, a factor is adopted
to suppress the unphysical contributions of the coupled-channel
effects in the high momentum region as done in the literature~\cite{Morel:2002vk,Silvestre-Brac:1991qqx,Zhong:2022cjx,Ortega:2016mms,Ortega:2016pgg,Yang:2022vdb,Yang:2021tvc}. As we know the vertices of the $^3P_0$ model
are only effective in the non-perturbative region, which reflect the ability of
$q\bar{q}$ creation in the vacuum. This ability will be suppressed in the
high momentum region due to the weak interactions between the valence quarks. Thus,
one need introduce a suppressed factor when extending the vertices
of $^3P_0$ model to the high momentum region.

Our main purposes of this work are as follows.
(i) To obtain a more comprehensive understanding of the charmonium
spectrum within the unquenched quark model. Here, we focus not only the
explanation of the masses but also the decay widths.
(ii) To systematically explore the magnitude of mass shifts of the higher charmonium
states up to the mass region of $\sim 5.0$ GeV due to the coupled-channel effects.
By this study, we expect to clarify whether the mass shifts have an obvious increasing trend from lower
levels to higher ones as that predicted with screened potentials.
(iii) To know whether the charmonium-like states could be accommodated by
the charmonium spectrum when including the unquenched coupled-channel effects.
It is also crucial for seeking out the genuine exotic states from the charmonium-like states.
(iv) To predict the masses and decay properties of the
missing higher charmonium states within a unquenched quark model.
We expect our predictions can provide useful references for the future
experimental observations.

The paper is organized as follows. In Sec.~\ref{Fram},
we give an introduction of the framework of the unquenched
quark model including coupled-channel effects.
In Sec.~\ref{Discussion}, the masses and strong decay
properties of the $S$-, $P$- and $D$-wave charmonium states up to mass region of
$\sim 5.0$ GeV are given. Then, we further give some
discussions based on the results. Finally, a summary is given in Sec.~\ref{SUM}.

\section{Framework}\label{Fram}

\subsection{Quark potential model}

The bare $c\bar{c}$ states are described within a quenched semirelativistic quark potential model.
In the model, the effective Hamiltonian is described by
\begin{equation}\label{H0}
 H_0=\sqrt{\mathbf{p}_1^2+m_{q}^2}+\sqrt{\mathbf{p}_2^2+m_{\bar{q}}^2}+V_0(r)+V_{sd}(r),
\end{equation}
where the first two terms stand for the kinetic energies for the quark and antiquark, respectively.
Their masses are labeled with $m_{q}$ and $m_{\bar{q}}$, respectively.
While their three momenta are labeled with $\mathbf{p}_1$ and $\mathbf{p}_2$, respectively.
$r$ is the distance between the quark and antiquark.
$V_0(r)$ is the well-known Cornell potential~\cite{Eichten:1978tg}
\begin{eqnarray}
V_0(r)=-\frac{4}{3}\frac{\alpha_s}{r}+br+C_0,
\end{eqnarray}
which includes the color Coulomb interaction and linear confinement, and zero point energy $C_0$.
The parameters $b$ and $\alpha_s$ denote the strength of the confinement and
strong coupling of the one-gluon-exchange potential, respectively.
The spin-dependent potential,
$V_{sd}(r)$, is adopted the widely used form~\cite{Godfrey:1985xj,Barnes:2005pb}
\begin{eqnarray}\label{Vsd}
\begin{aligned}
V_{sd}(r)&=\frac{32\pi \alpha_{s} \cdot \sigma^3 e^{-\sigma^2 r^2}}{9 \sqrt{\pi^3}  m_{q}m_{\bar{q}}} \mathbf{S_q} \cdot \mathbf{S_{\bar{q}}} \\
 &+\frac{4}{3}\frac{\alpha_s}{m_{q} m_{\bar{q}}}\frac{1}{r^3}\left(\frac{3\mathbf{S}_{q}\cdot \mathbf{r}\mathbf{S}_{{\bar{q}}}\cdot \mathbf{r}}{r^2}-\mathbf{S}_{q}\cdot\mathbf{S}_{{\bar{q}}}\right)+H_{LS}.
\end{aligned}
\end{eqnarray}
In the above equation, the first and second terms are the spin-spin contact hyperfine potential
and tensor potential, respectively. The third term is the spin-orbit interaction, which can be further decomposed into symmetric and antisymmetric terms~\cite{Godfrey:2004ya,Li:2019tbn}, i.e.,
\begin{eqnarray}\label{Lssyma}
H_{LS}=H_{sym}+H_{anti},
\end{eqnarray}
with
\begin{eqnarray}\label{Lssym}
H_{sym}=\frac{\mathbf{S_{+}\cdot L}}{2}\left[ \left(\frac{1}{2  m_q^2}+\frac{1}{2m_{\bar{q}}^2} \right) \left( \frac{4 \alpha_s}{3r^3}-\frac{b}{r}   \right)+\frac{8 \alpha_s}{3  m_q m_{\bar{q}} r^3} \right],
\end{eqnarray}
\begin{eqnarray}\label{Lsanti}
H_{anti}=\frac{\mathbf{S_{-}\cdot L}}{2}\left(\frac{1}{2  m_q^2}-\frac{1}{2m_{\bar{q}}^2} \right) \left( \frac{4 \alpha_s}{3r^3}-\frac{b}{r}   \right).
\end{eqnarray}
In these equations, $\mathbf{L}$ is the relative orbital angular momentum of the $q\bar{q}$
system; $\mathbf{S}_q$ and $\mathbf{S}_{\bar{q}}$ are the spins of the quark $q$ and antiquark $\bar{q}$, respectively, and $\mathbf{S}_{\pm}\equiv \mathbf{S}_q\pm \mathbf{S}_{\bar{q}}$. It should be mentioned that, for a $c\bar{c}$ state,
there are no contributions from the the antisymmetric term $H_{anti}$ due to the equal masses
for the quark and antiquark. The parameter set $\{ m_q, m_{\bar{q}}, b, \alpha_s, \sigma, C_0 $ in the
above potentials is determined by fitting the mass spectrum.
To solve the radial Schrodinger equation, the Gaussian expansion method~\cite{Hiyama:2003cu} is adopted in this work.

\begin{figure}
\centering \epsfxsize=5.6 cm \epsfbox{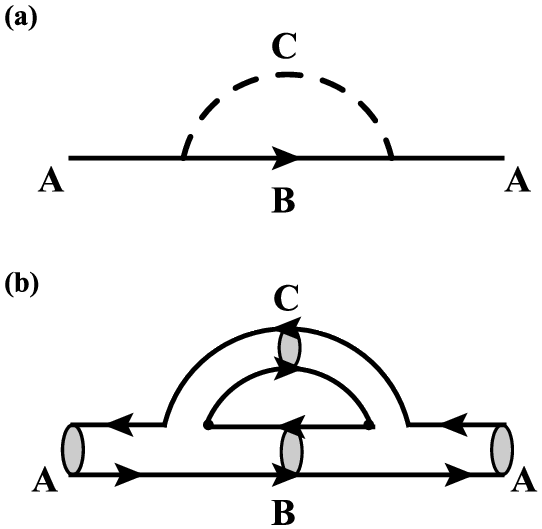} \vspace{-0.0 cm} \caption{Coupled-channel effects via the $BC$
hadronic loops for a bare meson state $|A \rangle$ shown at the hadronic level (a), and the quark level (b), respectively.}\label{CCEFsFig}
\end{figure}

\subsection{Coupled-channel effects}

There are creation and annihilation of $q\bar{q}$
pairs in vacuum  for a physical hadron state. Thus, a bare meson state $|A\rangle$ described within the quenched quark model
can further couple to two-hadron $BC$ continuum via hadronic loops, as shown in Fig.~\ref{CCEFsFig}.
Including such unquenched coupled-channel effects, in the simple coupled-channel model~\cite{Kalashnikova:2005ui}, the physical state is described by
\begin{equation}
| \Psi \rangle =c_A |A\rangle + \sum_{BC}\int c_{BC}(\mathbf{p})d^3\mathbf{p} |BC,\mathbf{p}\rangle ,
\end{equation}
where $\mathbf{p}=\mathbf{p}_B=-\mathbf{p}_C$ is the final two-hadron relative momentum in the initial hadron static system, $c_A$ and $c_{BC}(\mathbf{p})$ denote the probability amplitudes of the bare state $|A\rangle$ and $|BC,\mathbf{p}\rangle$ continuum, respectively.

The effective Hamiltonian of the physical state $| \Psi \rangle$ is given by
\begin{equation}\label{HHH}
H = H_0+H_c +H_I,
\end{equation}
where $H_0$ is the Hamiltonian for describing the bare state $|A\rangle$, which has been given by Eq.~(\ref{H0}).
$H_c$ is a Hamiltonian for describing the continuum state $|BC,\mathbf{p}\rangle$.
Neglecting the interactions between the $B$ and $C$ hadrons, the eigenenergy of
$|BC,\mathbf{p}\rangle$ is given by
\begin{equation}
E_{BC}=\sqrt{m_B^2+\mathbf{p}^2}+\sqrt{m_C^2+\mathbf{p}^2}.
\end{equation}
While $H_I$ is an effective Hamiltonian for describing the coupling of
the bare state$|A\rangle$ with the $|BC,\mathbf{p}\rangle$ continuum. In the present work,
this coupling is adopted from the
widely used $^3P_0$ model~\cite{Micu:1968mk,LeYaouanc:1972vsx,LeYaouanc:1973ldf} ,
which is expressed as
\begin{eqnarray}\label{T operator}
H_I&=&-3 \gamma \sqrt{96\pi } \sum_{m}\langle 1, m; 1,-m \mid 0,0\rangle \int d \mathbf{p}_{3} d \mathbf{p}_{4} \delta^{3}\left(\mathbf{p}_{3}+\mathbf{p}_{4}\right) \nonumber\\
&&\times \mathcal{Y}_{1}^{m}\left(\frac{\mathbf{p}_{3}-\mathbf{p}_{4}}{2}\right) \chi_{1-m}^{34} \phi_{0}^{34} \omega_{0}^{34} b_{3i}^{\dagger}\left(\mathbf{p}_{3}\right) d_{4j}^{\dagger}\left(\mathbf{p}_{4}\right),
\end{eqnarray}
where $\gamma$ is a dimensionless constant that denotes the strength of the quark-antiquark pair creation with momentum $\mathbf{p}_{3}$ and $\mathbf{p}_{4}$ from vacuum; $b_{3i}^{\dagger}\left(\mathbf{p}_{3}\right)$ and $d_{4j}^{\dagger}\left(\mathbf{p}_{4}\right)$ are the creation operators for the quark and antiquark, respectively; the subscripts, $i$ and $j$, are the SU(3)-color indices of the created quark and antiquark; $\phi_{0}^{34}=\left (u\bar{u}+d\bar{d}+s\bar{s}\right ) /\sqrt{3}$ and $\omega_{0}^{34}= \delta_{ij}/\sqrt{3}$ correspond to flavor and color singlets, respectively; $\chi_{1-m}^{34}$ is a spin triplet state; $\mathcal{Y}_{\ell m}(\mathbf{k}) \equiv|\mathbf{k}|^{\ell} Y_{\ell m}\left(\theta_{\mathbf{k}}, \phi_{\mathbf{k}}\right)$ is the $\ell$-th solid harmonic polynomial.

The Schr\"{o}dinger equation including coupled-channel effects can be expressed as
\begin{equation}\label{coupled-channel equation}
\begin{aligned}
&\left( \begin{matrix}H_0~~~~~~H_I
\\  H_I~~~~~~H_c \end{matrix}\right)
~\left( \begin{matrix} c_A |A\rangle
\\ \sum_{BC}\int c_{BC}(\mathbf{p})d^3\mathbf{p} |BC,\mathbf{p}\rangle  \end{matrix}\right)~~~~~~~~~~~~~~~~~~~~~~~~~~~~~~~\\
&~~~~~~~~~~~~~~~~~~~~~~~~~~~~~
=M \left( \begin{matrix} c_A |A\rangle
\\  \sum_{BC}\int c_{BC}(\mathbf{p}) d^3\mathbf{p}|BC,\mathbf{p}\rangle    \end{matrix}\right),
\end{aligned}
\end{equation}
where $M$ is the mass of the physical state $| \Psi \rangle$. It can be determined by the coupled-channel equation
\begin{equation}\label{M=MA+Delta M}
M=M_A+\Delta M(M).
\end{equation}
$M_A$ is the bare mass of $|A\rangle$ obtained from the potential model.
While $\Delta M(M)$ is the mass shift due to the coupled-channel effects, which
is given by
\begin{eqnarray}\label{ReDelta M}
\Delta M(M) &=&\mathrm{Re} \sum_{BC}\int_0^{\infty} \frac{|\langle BC,\mathbf{p} |H_I| A \rangle|^2}{(M-E_{BC})} p^2 dp d\Omega_p \nonumber \\
&=& \operatorname{Re} \sum_{B C} \int_{0}^{\infty} \frac{\overline{\left|\mathcal{M}_{A \rightarrow B C}(\mathbf{p})\right|^{2}}}{\left(M-E_{B C}\right)} p^{2} d p.
\end{eqnarray}

In principle, all $|BC\rangle$ hadronic loops should contribute a mass shift to the bare state $|A\rangle$.
However, it is unfeasible to calculate the self-energy function including an unlimited number of loops.
In our calculations, we included all OZI-allowed two-body hadronic channels with mass thresholds below the bare $|A\rangle$ states. Additionally, we account for channels with thresholds slightly (about $50$ MeV) higher than the bare mass. The contributions from the other far away virtual channels (whose mass thresholds are significant above the bare $|A\rangle$ states) are subtracted from the dispersion relation by redefining the bare mass with the once-subtracted method suggested in Ref.~\cite{Pennington:2007xr}.
In this approach, the mass shift can be evaluated with
\begin{equation}\label{ReDelta MM0}
\begin{aligned}
\Delta M(M)&= \operatorname{Re} \sum_{B C} \int_{0}^{\infty} \frac{\overline{\left|\mathcal{M}_{A \rightarrow B C}(\mathbf{p})\right|^{2}}}{\left(M-E_{B C}\right)} p^{2} d p \\
&-
\operatorname{Re} \sum_{B C} \int_{0}^{\infty} \frac{\overline{\left|\mathcal{M}_{A \rightarrow B C}(\mathbf{p})\right|^{2}}}{\left(M_0-E_{B C}\right)} p^{2} d p \\
&=\operatorname{Re} \sum_{BC} \int_{0}^{\infty}
\frac{\left(M_0-M\right)\overline{\left|\mathcal{M}_{A \rightarrow BC}(\mathbf{p})\right|^{2}}}{\left(M-E_{BC}\right)\left(M_0-E_{BC}\right)}
p^{2} d p,
\end{aligned}
\end{equation}
where $M_{0}$ represents the subtracted zero-point for the $c\bar{c}$ states.
It is chosen as $M_{0}=3097$ MeV, which is the mass of $J/\psi$.
This method has been successfully applied to study the coupled-channel effects
on heavy-light and charmonium states in the literature~\cite{Ni:HL,Zhou:2011sp,Duan:2021alw}.

The vertices given by the $^3P_0$ model are too hard at high momenta.
To suppress the unphysical contributions to the mass shift from the
higher momentum region, as done in the literature~\cite{Silvestre-Brac:1991qqx,Zhong:2022cjx,Ortega:2016mms,Ortega:2016pgg,Yang:2022vdb,Yang:2021tvc},
a suppressed factor $e^{-\mathbf{p}^2/(2 \Lambda^2)}$ is introduced in the strong transition amplitudes, i.e.,
\begin{eqnarray}\label{Form factor}
\mathcal{M}_{A \rightarrow B C}(\mathbf{p})
\to  \mathcal{M}_{A \rightarrow B C}(\mathbf{p})e^{-\frac{\mathbf{p}^2}{2 \Lambda^2}},
\end{eqnarray}
where, $\Lambda$ is a cut-off parameter. In this study, we adopt $\Lambda=780$ MeV to
consist with our recent study of the heavy-light meson spectrum
including unquenched coupled-channel effects~\cite{Ni:HL}.

By combining Eq.~(\ref{M=MA+Delta M}) and Eq.~(\ref{ReDelta MM0}), one can determine
the physical mass $M$ together with the mass shift $\Delta M$.
If the mass of the initial state $A$ is above the mass threshold
of final hadron states $B$ and $C$, a strong decay process $A\to BC$ will happen.
The partial decay width for the opened $BC$ channel is given by
\begin{eqnarray}\label{width}
\Gamma=2\pi \frac{|\mathbf{p}|E_BE_C}{M_A}\overline{\left|\mathcal{M}_{A \rightarrow B C}(\mathbf{p})\right|^{2}},
\end{eqnarray}
which is equal to two times of the imaginary part of
the self-energy of the $BC$ loop as shown in Fig.~\ref{CCEFsFig}.

\begin{table}
\begin{center}
\caption{The measured widths (in MeV) of the four well-established states, $\psi(3770)$, $\psi(4160)$,
$\psi(4040)$, and $\chi_{c2}(3930)$, compared with the fitted theoretical results.
The creation quark pair strength $\gamma=0.422$ is obtained by fitting the data
with $\chi^2/\mathrm{d.o.f}=5.4/(4-1)=1.8$. It should be noted that the errors of
some data are properly adjusted to obtain an overall successful fitting.}\label{chi-square}
\tabcolsep=0.305cm
\scalebox{1.0}{
\begin{tabular}{lccccccccc}
\hline
\hline
$n^{2S+1}L_J$  &Observed State   &$\Gamma_{\mathrm{exp}}$~\cite{ParticleDataGroup:2022pth} &$\Gamma_{\mathrm{Error}}$ &$\Gamma_{\mathrm{th}}$   \\
\hline
$1^{3}D_1$     &$\psi(3770)$       &$27.2\pm1.0$   &$5.0$  &$35.1$\\
$2^{3}D_1$     &$\psi(4160)$       &$70.0\pm10.0$  &$10.0$ &$61.8$\\
$3^{3}S_1$     &$\psi(4040)$       &$80.0\pm10.0$  &$10.0$ &$77.9$\\
$2^{3}P_2$     &$\chi_{c2}(3930)$  &$35.2\pm2.2$   &$10.0$ &$21.0$\\
\hline
\hline
\end{tabular}}
\end{center}
\end{table}

\subsection{Model parameters}

To consist with our previous work for the study of the charmed and charmed-strange mesons
in the potential model~\cite{Ni:2021pce}, the constituent masses for charmed, strange, and up/down quarks
are taken as $m_c=1.70$ GeV, $m_{s}=0.50$ GeV, and $m_{u,d}=0.40$ GeV, respectively.
The slope parameter $b$ for the linear confining potential is taken as $b=0.18~\mathrm{GeV}^2$.
The other three parameters $\{\alpha_s, \sigma, C_0\}$ are taken as
$\alpha_s=0.445$, $\sigma=1.20$ GeV, and $C_0=-495$ MeV, which are determined by fitting the masses of
the low-lying well established $1S$, $2S$, $1P$ and $1D$ states. To overcome the singular behavior of $1/r^3$ in the spin-dependent potentials, following the method of our previous works~\cite{li:2021hss,Deng:2016ktl,Deng:2016stx,Li:2019tbn,
Li:2020xzs,Ni:2021pce}, we introduce a cutoff distance $r_c$ in the calculations, i.e., let $1/r^3=1/r_c^3$ within a small range $r\in (0,r_c)$. The cutoff distance $r_c$ is taken as $r_c=0.166$ fm, which is determined by fitting the mass of $\chi_{c0}(1P)$,
since its mass is sensitive to the cutoff distance $r_c$.

When calculating the strong transition amplitudes,
one also need the masses and wave functions of the initial and final hadron states,
and the strength parameter for the quark pair creation of the $^3P_0$ model.
The masses of the well-established hadrons are taken from PDG~\cite{ParticleDataGroup:2022pth}.
For the unestablished charmonium states, their masses are taken from our predictions in the present work;
while for the unestablished $D$- and $D_s$-meson states, their masses are taken
from our previous predictions in Ref.~\cite{Ni:2021pce}.
The wave functions of the charmonium states and $D_{(s)}$-meson states are adopted
the numerical forms predicted in the present work and our previous work~\cite{Ni:2021pce},
respectively. The strength parameter for the quark pair creation
is taken as $\gamma=0.422$, which is determined by fitting the
widths of the well-established charmonium states $\psi(3770)$, $\psi(4040)$, $\psi(4160)$
and $\chi_{c2}(2P)$ with $\chi^2/\mathrm{d.o.f}=1.8$, as shown in Table~\ref{chi-square}.

\begin{figure*}
\centering \epsfxsize=14.2 cm \epsfbox{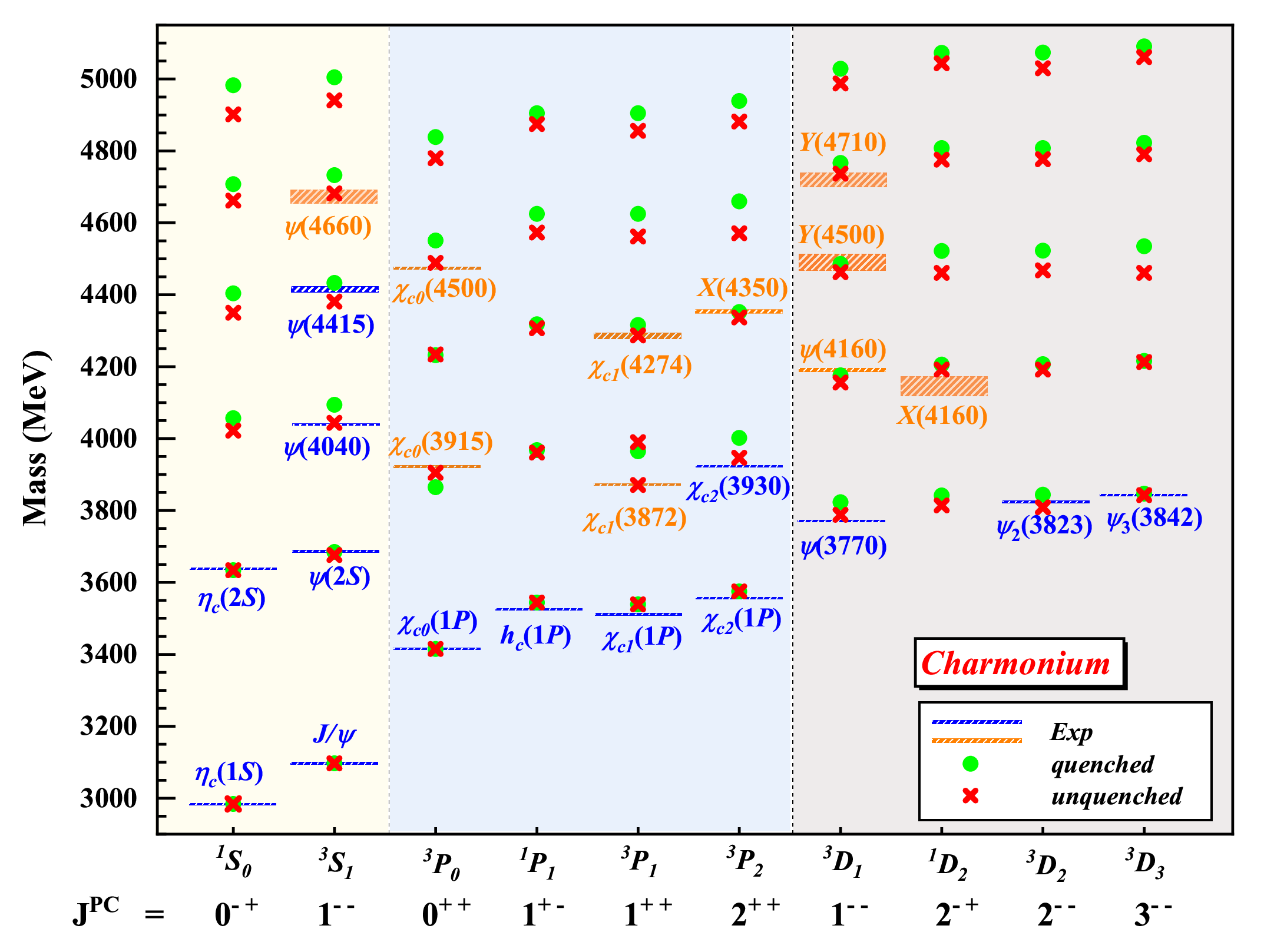} \vspace{-0.0 cm} \caption{Charmonium mass spectrum compared with the observations.
The well established states in the charmonium mass spectrum are labeled with blue color. The results obtained within
the unquenched and quenched quark models are labeled by a red cross and a solid green circle, respectively. }\label{spectrum}
\end{figure*}

\begin{table*}
\begin{center}
\caption{Charmonium mass spectrum described within both the quenched $(Q)$ and unquenched $(UQ)$ quark models.
The mass shifts ($\Delta M$) of the bare states due to the coupled-channel effects are given in the fourth column.
The unit of mass is MeV. For a comparison, the experimental data and some results from other works are also listed in the table.
The notation ``\ding{55}" stands for the case of no mass shift for a $c\bar{c}$ state.}
\label{ccbar mass spectrum}
\renewcommand\arraystretch{1.0}
\tabcolsep=0.160cm
\scalebox{1.0}{
\begin{tabular}{cccccccccccccccccccc}
\bottomrule[1.0pt]\bottomrule[1.0pt]
$n^{2S+1}L_{J}$
&   $J^{PC}$
&    $Q$
&    $\Delta M$
&    $UQ$
&    Exp \cite{ParticleDataGroup:2022pth}
&    LC \cite{Li:2009zu}
&   EFG \cite{Ebert:2011jc}
&   BGS \cite{Barnes:2005pb}
&   WCLM \cite{Wang:2019mhs}
&   DLGZ \cite{Deng:2016stx}
&   GI \cite{Godfrey:1985xj}\\
\bottomrule[0.5pt]	
$ 1^{1}S_{0}$      &     $0^{-+}$ &     $2984$     &  \ding{55}  &  $2984$      &     $2984$      &     $2979$ &     $2981$ &     $2982$   &     $2981$ &     $2983$ &     $2970$ \tabularnewline		
$ 1^{3}S_{1}$      &     $1^{--}$ &     $3097$     &  \ding{55}  &  $3097$      &     $3097$      &     $3097$ &     $3096$ &     $3090$   &     $3096$ &     $3097$ &     $3100$ \tabularnewline				
$ 2^{1}S_{0}$      &     $0^{-+}$ &     $3634$     &  \ding{55}  &  $3634$      &     $3639$      &     $3623$ &     $3635$ &     $3630$   &     $3642$ &     $3635$ &     $3620$ \tabularnewline				
$ 2^{3}S_{1}$      &     $1^{--}$ &     $3685$     &  $-9$ &  $3676$      &     $3686$      &     $3673$ &     $3685$ &     $3672$   &     $3683$ &     $3679$ &     $3680$ \tabularnewline			
$ 3^{1}S_{0}$      &     $0^{-+}$ &     $4057$     &  $-35$      &  $4022$      &     $...$       &     $3991$ &     $3989$ &     $4043$   &     $4013$ &     $4048$ &     $4060$ \tabularnewline	
$ 3^{3}S_{1}$      &     $1^{--}$ &     $4094$     &  $-50$      &  $4044$      &     $4039$      &     $4022$ &     $4039$ &     $4072$   &     $4035$ &     $4078$ &     $4100$ \tabularnewline		
$ 4^{1}S_{0}$      &     $0^{-+}$ &     $4404$     &  $-54$      &  $4350$      &     $...$       &     $4250$ &     $4401$ &     $4384$   &     $4260$ &     $4388$ &     $...$ \tabularnewline		
$ 4^{3}S_{1}$      &     $1^{--}$ &     $4433$     &  $-52$      &  $4381$      &     $4421$  &     $4273$ &     $4427$ &     $4406$   &     $4274$ &     $4412$ &     $4450$ \tabularnewline			
$ 5^{1}S_{0}$      &     $0^{-+}$ &     $4708$     &  $-46$      &  $4662$      &     $...$       &     $4446$ &     $4811$ &     $...$    &     $4433$ &     $4690$ &     $...$ \tabularnewline
$ 5^{3}S_{1}$      &     $1^{--}$ &     $4733$     &  $-51$      &  $4682$      &     $4652$~\cite{Belle:2014wyt}       &     $4463$ &     $4837$ &     $...$    &     $4443$ &     $4711$ &     $...$ \tabularnewline
$ 6^{1}S_{0}$      &     $0^{-+}$ &     $4983$     &  $-81$      &  $4902$      &     $...$       &     $4595$ &     $5155$ &     $...$    &     $...$  &     $...$  &     $...$ \tabularnewline
$ 6^{3}S_{1}$      &     $1^{--}$ &     $5005$     &  $-64$      &  $4941$      &     $...$       &     $4608$ &     $5167$ &     $...$    &     $...$  &     $...$  &     $...$ \tabularnewline
\bottomrule[0.3pt]
$ 1^{3}P_{0}$      &     $0^{++}$ &     $3415$     &  \ding{55}  &  $3415$      &     $3415$      &     $3433$ &     $3413$ &     $3424$   &     $3464$ &     $3415$ &     $3440$ \tabularnewline				
$ 1^{1}P_{1}$      &     $1^{+-}$ &     $3544$     &  \ding{55}  &  $3544$      &     $3525$      &     $3519$ &     $3525$ &     $3516$   &     $3538$ &     $3522$ &     $3520$ \tabularnewline		
$ 1^{3}P_{1}$      &     $1^{++}$ &     $3539$     &  \ding{55}  &  $3539$      &     $3511$      &     $3510$ &     $3511$ &     $3505$   &     $3530$ &     $3516$ &     $3510$ \tabularnewline
$ 1^{3}P_{2}$      &     $2^{++}$ &     $3575$     &  \ding{55}  &  $3575$      &     $3556$      &     $3554$ &     $3555$ &     $3556$   &     $3571$ &     $3552$ &     $3550$ \tabularnewline
\bottomrule[0.3pt]
$ 2^{3}P_{0}$      &     $0^{++}$ &     $3865$     &  $+40$      &  $3905$      &     $3922$      &     $3842$ &     $3870$ &     $3852$   &     $3896$ &     $3869$ &     $3920$ \tabularnewline			
$ 2^{1}P_{1}$      &     $1^{+-}$ &     $3968$     &  $-7$       &  $3961$      &     $...$       &     $3908$ &     $3926$ &     $3934$   &     $3933$ &     $3940$ &     $3960$ \tabularnewline
$ 2^{3}P_{1}$      &     $1^{++}$ &     $3965$     &  $-94/+25$   &  $3871/3990$ &     $3872/...$      &     $3901$ &     $3906$ &     $3925$   &     $3929$ &     $3937$ &     $3950$ \tabularnewline
$ 2^{3}P_{2}$      &     $2^{++}$ &     $4002$     &  $-55$      &  $3947$      &     $3923$      &     $3937$ &     $3949$ &     $3972$   &     $3952$ &     $3967$ &     $3980$ \tabularnewline				
\bottomrule[0.3pt]
$ 3^{3}P_{0}$      &     $0^{++}$ &     $4232$     &  $+2$       &  $4234$      &     $...$       &     $4131$ &     $4301$ &     $4202$   &  $4177$   &  $4230$ &     $...$ \tabularnewline			
$ 3^{1}P_{1}$      &     $1^{+-}$ &     $4318$     &  $-11$      &  $4307$      &     $...$       &     $4184$ &     $4337$ &     $4279$   &  $4200$   &  $4285$ &     $...$ \tabularnewline
$ 3^{3}P_{1}$      &     $1^{++}$ &     $4316$     &  $-29$      &  $4287$      &     $4286$& $4178$ &     $4319$ &     $4271$   &  $4197$   &  $4284$ &     $...$ \tabularnewline
$ 3^{3}P_{2}$      &     $2^{++}$ &     $4352$     &  $-15$      &  $4337$      &     $4351$~\cite{Belle:2009rkh}      &     $4208$ &     $4354$ &     $4317$   &  $4213$   &  $4310$ &     $...$ \tabularnewline			
\bottomrule[0.3pt]
$ 4^{3}P_{0}$      &     $0^{++}$ &     $4551$     &  $-62$      &  $4489$      &     $4474$        &$...$ &     $4698$ &     $...$   &   $4374$ &     $...$ &     $...$ \tabularnewline			
$ 4^{1}P_{1}$      &     $1^{+-}$ &     $4625$     &  $-52$      &  $4573$      &     $...$             &$...$ &     $4744$ &     $...$   &   $4389$ &     $...$ &     $...$ \tabularnewline
$ 4^{3}P_{1}$      &     $1^{++}$ &     $4625$     &  $-63$      &  $4562$      &     $...$             &$...$ &     $4728$ &     $...$   &   $4387$ &     $...$ &     $...$ \tabularnewline
$ 4^{3}P_{2}$      &     $2^{++}$ &     $4660$     &  $-89$      &  $4571$      &     $...$             &$...$ &     $4763$ &     $...$   &   $4398$ &     $...$ &     $...$ \tabularnewline					
\bottomrule[0.3pt]
$ 5^{3}P_{0}$      &     $0^{++}$ &     $4839$     &  $-59$      &  $4780$      &     $...$             &$...$ &     $...$ &       $...$   &     $...$ &     $...$ &     $...$ \tabularnewline			
$ 5^{1}P_{1}$      &     $1^{+-}$ &     $4905$     &  $-30$      &  $4875$      &     $...$             &$...$ &     $...$ &       $...$   &     $...$ &     $...$ &     $...$ \tabularnewline
$ 5^{3}P_{1}$      &     $1^{++}$ &     $4905$     &  $-49$      &  $4856$      &     $...$             &$...$ &     $...$ &       $...$   &     $...$ &     $...$ &     $...$ \tabularnewline
$ 5^{3}P_{2}$      &     $2^{++}$ &     $4939$     &  $-57$      &  $4882$      &     $...$             &$...$ &     $...$ &       $...$   &     $...$ &     $...$ &     $...$ \tabularnewline	
\bottomrule[0.3pt]
$ 1^{3}D_{1}$      &     $1^{--}$ &     $3823$     &  $-35$      &  $3788$      &     $3774$      &     $3787$ &     $3783$ &     $3785$   &     $3830$ &     $3787$ &     $3820$ \tabularnewline
$ 1^{1}D_{2}$      &     $2^{-+}$ &     $3842$     &  $-28$   &  $3814$      &     $...$       &     $3796$ &     $3807$ &     $3799$   &     $3848$ &     $3806$ &     $3840$ \tabularnewline	
$ 1^{3}D_{2}$      &     $2^{--}$ &     $3844$     &  $-35$   &  $3809$      &     $3824$      &     $3798$ &     $3795$ &     $3800$   &     $3848$ &     $3807$ &     $3840$ \tabularnewline
$ 1^{3}D_{3}$      &     $3^{--}$ &     $3847$     &  $-4$       &  $3843$      &     $3843$      &     $3799$ &     $3813$ &     $3806$   &     $3859$ &     $3811$ &     $3850$ \tabularnewline		
\bottomrule[0.3pt]
$ 2^{3}D_{1}$      &     $1^{--}$ &     $4176$     &  $-20$      &  $4156$      &     $4191$     &     $4089$ &     $4150$ &     $4142$   &     $4125$ &     $4144$ &     $4190$ \tabularnewline       				
$ 2^{1}D_{2}$      &     $2^{-+}$ &     $4206$     &  $-14$      &  $4192$      &     $4146$~\cite{LHCb:2021uow}       &     $4099$ &     $4196$ &     $4158$   &     $4137$ &     $4164$ &     $4210$ \tabularnewline       				
$ 2^{3}D_{2}$      &     $2^{--}$ &     $4207$     &  $-15$      &  $4192$      &     $...$       &     $4100$ &     $4190$ &     $4158$   &     $4137$ &     $4165$ &     $4210$ \tabularnewline       				
$ 2^{3}D_{3}$      &     $3^{--}$ &     $4216$     &  $-3$       &  $4213$      &     $...$       &     $4103$ &     $4220$ &     $4167$   &     $4144$ &     $4172$ &     $4220$ \tabularnewline       				
\bottomrule[0.3pt]
$ 3^{3}D_{1}$      &     $1^{--}$ &     $4486$     &  $-23$      &  $4463$      &     $4485$~\cite{BESIII:2022joj}       &     $4317$ &     $4507$ &     $...$    &     $4334$ &     $4456$ &     $4520$ \tabularnewline				
$ 3^{1}D_{2}$      &     $2^{-+}$ &     $4522$     &  $-61$      &  $4461$      &     $...$       &     $4326$ &     $4549$ &     $...$    &     $4343$ &     $4478$ &     $...$ \tabularnewline  	 			
$ 3^{3}D_{2}$      &     $2^{--}$ &     $4523$     &  $-55$      &  $4468$      &     $...$       &     $4327$ &     $4544$ &     $...$    &     $4343$ &     $4478$ &     $...$ \tabularnewline
$ 3^{3}D_{3}$      &     $3^{--}$ &     $4535$     &  $-74$      &  $4461$      &     $...$       &     $4331$ &     $4574$ &     $...$    &     $4348$ &     $4486$ &     $...$ \tabularnewline
\bottomrule[0.3pt]
$ 4^{3}D_{1}$      &     $1^{--}$ &     $4767$     &  $-30$      &  $4737$      &     $\sim4710$~\cite{BESIII:2023wqy,BESIII:2022kcv}       &     $...$  &     $4857$ &     $...$    &     $4484$ &     $...$  &     $...$ \tabularnewline			
$ 4^{1}D_{2}$      &     $2^{-+}$ &     $4808$     &  $-32$      &  $4776$      &     $...$       &     $...$  &     $4898$ &     $...$    &     $4490$ &     $...$  &     $...$\tabularnewline				
$ 4^{3}D_{2}$      &     $2^{--}$ &     $4808$     &  $-31$      &  $4777$      &     $...$       &     $...$  &     $4896$ &     $...$    &     $4490$ &     $...$  &     $...$ \tabularnewline
$ 4^{3}D_{3}$      &     $3^{--}$ &     $4823$     &  $-32$      &  $4791$      &     $...$       &     $...$  &     $4920$ &     $...$    &     $4494$ &     $...$  &     $...$\tabularnewline				
\bottomrule[0.3pt]
$ 5^{3}D_{1}$      &     $1^{--}$ &     $5029$     &  $-41$      &  $4988$      &     $...$       &     $...$  &     $...$  &     $...$    &     $...$  &     $...$  &     $...$\tabularnewline				
$ 5^{1}D_{2}$      &     $2^{-+}$ &     $5073$     &  $-29$      &  $5044$      &     $...$       &     $...$  &     $...$  &     $...$    &     $...$  &     $...$  &     $...$\tabularnewline				
$ 5^{3}D_{2}$      &     $2^{--}$ &     $5074$     &  $-44$      &  $5030$      &     $...$       &     $...$  &     $...$  &     $...$    &     $...$  &     $...$  &     $...$\tabularnewline
$ 5^{3}D_{3}$      &     $3^{--}$ &     $5091$     &  $-30$      &  $5061$      &     $...$       &     $...$  &     $...$  &     $...$    &     $...$  &     $...$  &     $...$\tabularnewline
\bottomrule[1.0pt]\bottomrule[1.0pt]
\end{tabular}}
\end{center}
\end{table*}

	\begin{table}
		\begin{center}	
			\caption{The mass shifts and strong decay widths for the $3S$- and $4S$-wave charmonium states (in MeV).
The ``\ding{55}'' is labeled the strong decay channels which are
not open or forbidden. These channels are considered to have no contributions to
the mass shifts and decay widths. The ``$...$'' stands for the predicted
values are negligibly small.}
			\label{3S4S}
			\tabcolsep=0.22cm
			\renewcommand\arraystretch{0.65}
			\scalebox{1.0}{
				\begin{tabular}{cccccccccccccccc}
					\bottomrule[1.0pt]\bottomrule[1.0pt]
					&\multicolumn{2}{c}{$3^1S_0$}  &  &\multicolumn{2}{c}{$3^3S_1$}
					\\
					\cline{2-3}\cline{5-6}
					&$\Delta M$   &$\Gamma$     &   &$\Delta M$  &$\Gamma$         \\
					Channel       &$[4057]$     &$[4022]$    &   &$[4094]$    &$as~\psi(4040)$    \\
					\bottomrule[0.4pt]
					$DD$               &\ding{55}   &\ding{55}     &   &$-1.24$     &$0.96$      \\
					$D_sD_s$           &\ding{55}   &\ding{55}     &   &$+0.07$     &$2.10$      \\
					$DD^*$             &$+17.06$    &$52.75$       &   &$+9.26$     &$17.00$     \\
					$D_sD_s^*$         &$-6.15$     &\ding{55}     &   &$-3.89$     &\ding{55}   \\
					$D^*D^*$           &$-45.56$    &$9.56$        &   &$-54.39$    &$57.88$     \\
					$D_s^*D_s^*$       &\ding{55}   &\ding{55}     &   &\ding{55}   &\ding{55}   \\
					\emph{Total}       &$-34.65$    &$62.31$       &   &$-50.19$    &$77.94$     \\
					$M_{th},\Gamma_{th}$    &\multicolumn{2}{c}{$\bm{4022,62.31}$}    &   &\multicolumn{2}{c}{$\bm{4044,77.94}$}   \\
					$M_{exp},\Gamma_{exp}$  &\multicolumn{2}{c}{$\bm{-,-}$}           &    &\multicolumn{2}{c}{$\bm{4039,80\pm10}$} \\
					\bottomrule[0.7pt]\bottomrule[0.7pt]
					&\multicolumn{2}{c}{$4^1S_0$}  &  &\multicolumn{2}{c}{$4^3S_1$}   \\
					\cline{2-3}\cline{5-6}
					&$\Delta M$      &$\Gamma$        &
					&$\Delta M$      &$\Gamma$        \\
					Channel          &$[4404]$       &$[4350]$           &
					&$[4433]$       &$as~\psi(4415)$   \\
					\bottomrule[0.4pt]
					$DD$                         &\ding{55}   &\ding{55}    &   &$-0.31$     &$0.70$       \\
					$D_sD_s$                     &\ding{55}   &\ding{55}    &   &$-0.05$     &$...$        \\
					$DD^*$                       &$+1.40$     &$1.11$       &   &$-0.07$     &$0.13$       \\
					$D_sD_s^*$                   &$-0.68$     &$2.26$       &   &$-0.12$     &$1.15$       \\
					$D^*D^*$                     &$+1.08$     &$11.69$      &   &$+2.74$     &$5.77$       \\
					$D_s^*D_s^*$                 &$+0.24$     &$0.43$       &   &$-0.89$     &$1.27$       \\
					$DD_{0}(2550)$               &\ding{55}   &\ding{55}    &   &$-3.50$     &$7.46$       \\
					$DD_{0}^{*}(2300)$           &$+2.62$     &$34.05$      &   &\ding{55}   &\ding{55}   \\
					$D_{s}D_{s0}^{*}(2317)$      &$-6.29$     &$3.36$       &   &\ding{55}   &\ding{55}   \\
					$D^{*}D_{0}^{*}(2300)$       &\ding{55}   &\ding{55}    &   &$-4.41$     &$20.99$      \\
					$D_s^{*}D_{s0}^{*}(2317)$    &\ding{55}   &\ding{55}    &   &$-1.33$     &\ding{55}   \\
					$DD_{2}^{*}(2460)$           &$-25.00$    &$7.65$       &   &$-9.37$     &$18.40$      \\
					$DD_{1}(2430)$               &\ding{55}   &\ding{55}    &   &$-9.89$     &$17.48$      \\
					$D_{s}D_{s1}(2460)$          &\ding{55}   &\ding{55}    &   &$-3.86$     &\ding{55}    \\
					$DD_{1}(2420)$               &\ding{55}   &\ding{55}    &   &$-2.98$     &$15.39$       \\
					$D^{*}D_{1}(2430)$           &$-12.14$    &\ding{55}    &   &$-7.57$     &\ding{55}     \\
					$D^{*}D_{1}(2420)$           &$-15.28$    &\ding{55}    &   &$-10.58$    &\ding{55}     \\
					\emph{Total}                 &$-54.05$    &$60.55$      &   &$-52.19$    &$88.74$         \\
					$M_{th},\Gamma_{th}$    &\multicolumn{2}{c}{$\bm{4350,60.55}$}  &  &\multicolumn{2}{c}{$\bm{4381,88.74}$}        \\
					$M_{exp},\Gamma_{exp}$  &\multicolumn{2}{c}{$\bm{-,-}$}         &  &\multicolumn{2}{c}{$\bm{4421\pm4,62\pm20}$} \\
					\bottomrule[1.0pt]\bottomrule[1.0pt]
			\end{tabular}}
		\end{center}
	\end{table}

\section{Result and Discussion}\label{Discussion}

The masses for the $S$-, $P$- and $D$-wave charmonium states up to mass region of $\sim 5.0$ GeV
obtained within the unquenched quark models are listed in Table~\ref{ccbar mass spectrum}.
For a comparison, the results of the quenched scenario, the experimental data, and some predictions from other works
are also presented in the same table.
For clarity, our theoretical mass spectra compared with the data
are plot Fig.~\ref{spectrum} as well. Furthermore, the mass shift and partial 
strong decay width contributed by each channels for the charmonium states are given in Tables~\ref{3S4S}-\ref{51D253D2}.

\subsection{$S$-wave states}\label{S-wave}

The mass spectrum up to $6S$-wave states is given in Table~\ref{ccbar mass spectrum} and
also shown in Fig.~\ref{spectrum}. The strong decay properties for the
higher $S$-wave states are given in Tables~\ref{3S4S}-\ref{6S}.
The masses for the low-lying $S$-wave $c\bar{c}$ states $\eta_c$, $\eta_c(2S)$, $J/\psi(1S)$ and $\psi(2S)$ can be well described
within the quark model. For the high-lying $3,4,5,6S$-wave states, from Table~\ref{ccbar mass spectrum} one can see that the unquenched coupled-channel effects of intermediate hadron loops have sizeable corrections to the masses of the bare states,
the mass shifts are predicted to be in the range of $\sim(-80,-30)$ MeV.

\subsubsection{$\psi(4040)$}

The well established vector state $\psi(4040)$, as
the assignment of $\psi(3S)$, from Table~\ref{3S4S}, one can see that both its mass and width
are consistent with the predictions. The predicted partial width ratio between $DD$ and $DD^*$ channels,
\begin{eqnarray}
\frac{\Gamma(DD)}{\Gamma(DD^*)}\simeq 0.06,
\end{eqnarray}
is close to the lower limit of the data $0.24\pm 0.05\pm0.12$ measured by the
\emph{BABAR} collaboration~\cite{BaBar:2009elc}. However, the predicted ratio
\begin{eqnarray}
\frac{\Gamma(D^*D^*)}{\Gamma(DD^*)}\simeq 3.4,
\end{eqnarray}
is inconsistent with the measured value $0.18\pm 0.14\pm0.03$ by \emph{BABAR}.
It should be mentioned that the $D^*D^*$ as the dominant channel of $\psi(4040)$ is supported by
the measurements of Belle~\cite{Belle:2006hvs}.

\subsubsection{$\psi(4415)$}

Considering $\psi(4415)$ as the $\psi(4S)$ assignment,
from Table~\ref{3S4S}, one can see that both the mass and width are consistent with the quark
model expectations. The predicted partial width ratios,
\begin{eqnarray}
\frac{\Gamma(DD)}{\Gamma(D^*D^*)}\simeq 0.12,\ \ \frac{\Gamma(DD^*)}{\Gamma(D^*D^*)}\simeq 0.02,
\end{eqnarray}
are consistent with the data $0.14\pm 0.12\pm0.03$ and $0.17\pm 0.25\pm0.03$ measured by the
\emph{BABAR} collaboration~\cite{BaBar:2009elc}, respectively, within uncertainties. Recently, the $\psi(4415)$ was observed in the $D_s^{*+}D_s^{*-}$ final state by
the BESIII collaboration~\cite{BESIII:2023wsc}. As the $\psi(4S)$ assignment, we find that the
branching fraction of $\psi(4415)$ into $D_s^{*+}D_s^{*-}$ is predicted to
be $\sim 1.5\%$ (see Table~\ref{3S4S}), which is similar to that into $\bar{D}_sD_s^{*}+c.c.$.
Thus, $\psi(4415)$ should be observed in the $\bar{D}_sD_s^{*}+c.c.$ final states as well.

It should be mentioned that the quark model classification
for $\psi(4415)$ still bears some controversies.
In the literature~\cite{Li:2009zu,Wang:2023zxj,Wang:2022jxj}, $\psi(4415)$ is suggested to be a
$5S$ state or a $5S$-$4D$ mixing state based on the screening potential model.
The screening effect is considered to be partly equivalent to the coupled-channel effect.
The magnitude of mass shifts due to the coupled-channel effects estimated within the screening
potential model reaches up to a fairly large value $\sim 200-300$ MeV,
which is about a factor of $\sim5-10$ larger than our estimations with
a more comprehensive consideration of the coupled-channel effects (see Table~\ref{3S4S}).
In Ref.~\cite{Gui:2018rvv}, the strong decays of $\psi(4415)$ as the $\psi(5S)$ assignment
based on the screening potential model have been studied by our group, it is found that the
decay width $\Gamma\simeq 8.4$ MeV is notably smaller than the experimental value
$\Gamma_{exp}\simeq 62\pm 20$ MeV. Thus, $\psi(4415)$ as the $\psi(5S)$ assignment
should be excluded according to our study within the unquenched quark model.

\subsubsection{$\psi(5S)$ and $\psi(6S)$}

The center-of-mass energies of BESIII experiments have been
extended to 4.95 GeV, which provide good opportunities for
establishing high vector charmonium states $\psi(5S)$ and $\psi(6S)$.
Our predictions about their masses and
decay properties have been listed in Tables~\ref{5S} and~\ref{6S}.

The $\psi(5S)$ may favor the $\psi(4660)$ assignment listed in RPP~\cite{ParticleDataGroup:2022pth}.
The predicted mass and width of $\psi(5S)$, $M=4682$ MeV and $\Gamma=76$ MeV,
are in good agreement with the data $M_{exp}=4652\pm 21$ MeV and $\Gamma_{exp}=68\pm 16$ MeV
measured at Belle~\cite{Belle:2014wyt}.
Recently, a new vector resonance $Y(4710)$ was observed in $e^+e^-\to K\bar{K}J/\psi$ at BESIII
~\cite{BESIII:2023wqy,BESIII:2022kcv}, the measured mass and width are close
to those of $\psi(4660)$. The $Y(4710)$
and $\psi(4660)$ may correspond to the same state. Further measurements
of the main decay channels listed in Table~\ref{5S}, such as $D^*D^*_1(2600)$ and $DD_1(2430)$,
may be useful to establish the $\psi(5S)$ state.

Finally, it is should
be mentioned that $\psi(4660)$ is also suggested to be assigned as
the $\psi(5S)$ state in Refs.~\cite{Ding:2007rg,Gui:2018rvv,Segovia:2008zz,Zhao:2023hxc}
based on some quenched quark model studies.


\subsection{$P$-wave states}\label{P-wave}

The mass spectrum up to $5P$-wave states is given in Table~\ref{ccbar mass spectrum} and
also shown in Fig.~\ref{spectrum}. The strong decay properties for the
$P$-wave states are given in Tables~\ref{2P3P}-\ref{51P153P1}.
All of the four low-lying $1P$-wave states, $h_c(1P)$ and $\chi_{c0,1,2}(1P)$, have been well established.
However, the situation become complicated and confusing when toward establishing the higher $P$-wave states.
The unquenched coupled-channel effects play crucial roles for understanding the nature
of $\chi_{cJ}(2P)$ states. The coupled-channel effects on the $3P$-wave states
are small. For the $4P$- and $5P$-wave states, the unquenched coupled-channel
effects systematically lower the mass spectrum, the mass shifts are predicted
to be in the range of $\sim(-90,-30)$ MeV.

\begin{figure}
	\centering \epsfxsize=7.6 cm \epsfbox{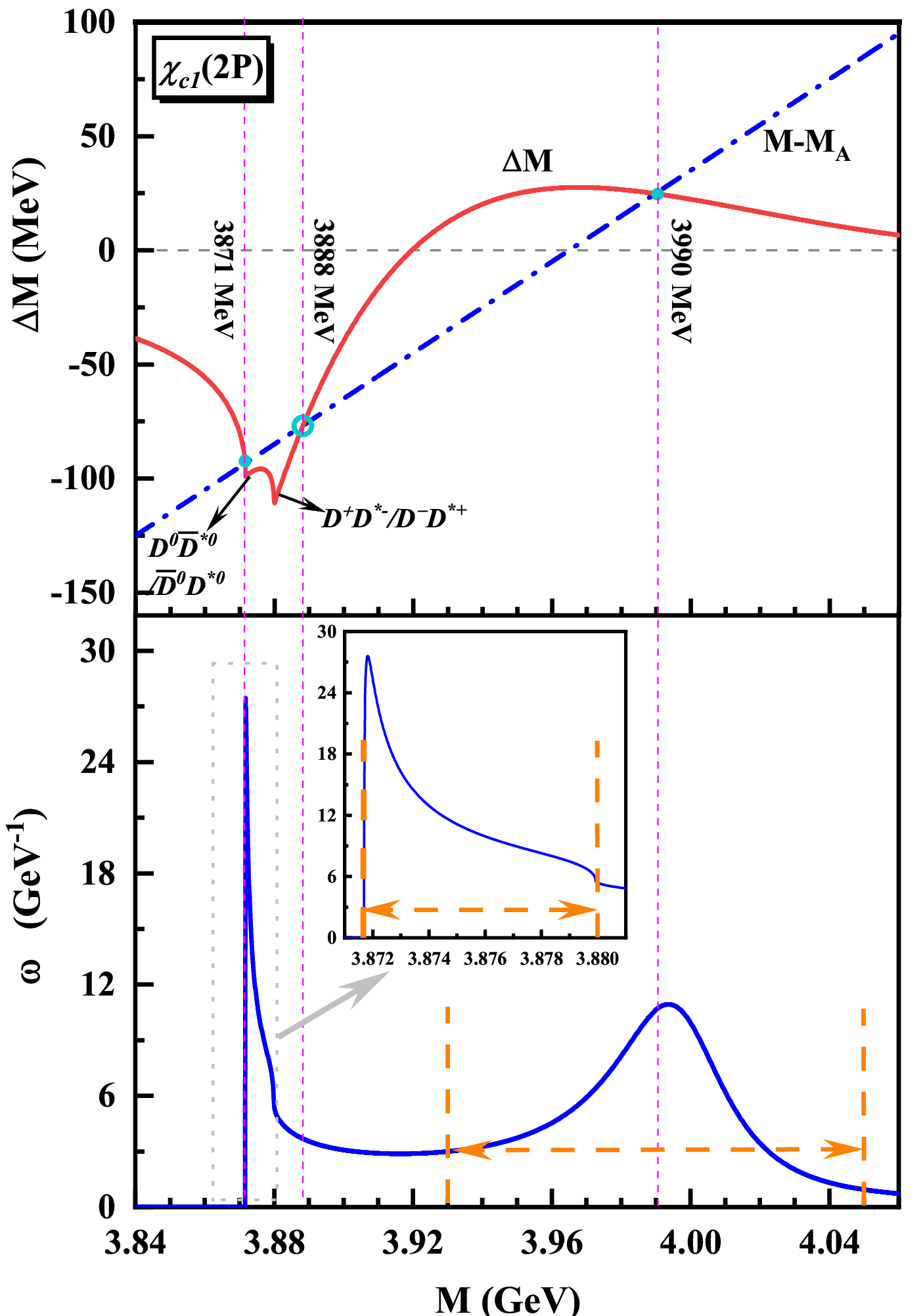} \vspace{-0.0 cm} \caption{The mass shift $\Delta M (M)$ (upper panel)
and spectral density function $\omega (M)$ (lower panel) for the $\chi_{c1}(2P)$ including the coupled-channel effect of $DD^{*}$.
Three solutions are found by solving the coupled-channel equation (\ref{M=MA+Delta M}), i.e.,
the intersection points of the two lines of $\Delta M (M)$ and $M-M_A$. The first and third solution
with masses of 3871 MeV and 3990 MeV correspond to the narrow and broad structures shown in the
spectral density function, respectively. The middle solution with a mass of 3888 MeV is unphysical. }\label{Chic12P}
\end{figure}

\begin{table}
\begin{center}
\caption{The components of the two $1^{++}$ resonances obtained by the $\chi_{c1}(2P)$ state
coupling to the $DD^*$ channel.}\label{Chic12Pcomponents}
		\tabcolsep=0.62cm
\scalebox{1.0}{
\begin{tabular}{ccccccccccccccccccccccccccccccccccccc}\bottomrule[1.0pt]\bottomrule[1.0pt]
			   Physical Mass           &$c\bar{c}$ core    &$DD^{*}$  \\
\bottomrule[0.8pt]
  $ 3871$ MeV    &$9.68\%$           &$90.32\%$  \\
  $3990$ MeV    &$61.11\%$          &$38.89\%$  \\
\bottomrule[1.0pt]\bottomrule[1.0pt]
\end{tabular}}
\end{center}
\end{table}


\subsubsection{$2P$-wave states}

For the $2P$-wave states, there are several candidates,
$\chi_{c1}(3872)$, $\chi_{c0}(3860)$, $\chi_{c0}(3930)$, $X(3915)$, and $\chi_{c2}(3930)$,
from experiments~\cite{ParticleDataGroup:2022pth}. However, about their assignments
one will face several problems: (i) Can $\chi_{c1}(3872)$ be assigned to $\chi_{c1}(2P)$
indeed? (ii) Is the mass gap between $\chi_{c0}(2P)$ and $\chi_{c2}(2P)$ small or large?
(iii) Is the width of $\chi_{c0}(2P)$ broad or narrow?
The unquenched quark model including couple-channel effects may shed light
on these puzzles.

First, let's focus on the $\chi_{c0}(2P)$ state. Within the unquenched quark model, the mass of $\chi_{c0}(2P)$
is predicted to be $M=3905$ MeV. The coupled-channel effects due to the $DD$-loop have a
significant correction to bare mass of $\chi_{c0}(2P)$.
By solving the coupled-channel equation, it is interesting to find that there is a positive mass
shift $\Delta M\simeq +40$ MeV. The positive mass
shift was also found by the Lanzhou Group~\cite{Duan:2020tsx}.
The $\chi_{c0}(2P)$ should be a narrow state with a width
of $\Gamma\simeq 16.8$ MeV, which is nearly saturated by the $DD$ channel.
Our prediction is consistent with that in Refs.~\cite{Duan:2020tsx,Liu:2009fe}.
Recently, in the $D^+D^-$ final state the LHCb collaboration observed a new $0^{++}$ charmonium resonance
$\chi_{c0}(3930)$ with mass $M_{exp}=3923.8\pm 1.9$ MeV and width
$\Gamma_{exp}=17.4\pm 5.9$ MeV~\cite{LHCb:2020pxc}. Comparing the observations with our predictions,
we find that $\chi_{c0}(3930)$ perfectly favors the $\chi_{c0}(2P)$ assignment.
The $\chi_{c0}(3930)$ resonance may correspond to $X(3915)$ observed in
the $\omega J/\psi$ final state by Belle~\cite{Belle:2004lle,Belle:2009and} and \emph{BABAR}~\cite{BaBar:2012nxg,BaBar:2007vxr}.

In Ref.~\cite{Guo:2012tv}, the authors analyzed the Belle and \emph{BABAR} data
of $\gamma\gamma \to D\bar{D}$. From the data, a broad resonance with mass
$3837.6\pm 11.5$ MeV and width $221\pm 19$ MeV was extracted. They claimed that
this broad resonance should correspond to $\chi_{c0}(2P)$, rather
than the narrow $X(3915)$ resonance. Soon after the work
was published, the Belle collaboration reported the observation
of a $\chi_{c0}(3860)$ state with a mass of $M_{exp}=3862^{+66}_{-45}$ MeV and
a width of $\Gamma_{exp}=201^{+242}_{-149}$ MeV in the process $e^+e^-\to J/\psi D\bar{D}$~\cite{Belle:2017egg}.
However, based our present study, the $\chi_{c0}(3860)$ resonance disfavors
the $\chi_{c0}(2P)$ assignment. The broad width is out of theoretical expectation,
although the mass seems to be consistent with the unquenched quark model predictions.
Our conclusion is consistent with that in Refs.~\cite{Duan:2020tsx,Gui:2018rvv}.
It should be pointed out that the $\chi_{c0}(3860)$ is not seen
in a recent observation of $B\to D^+D^-K^+$ at LHCb~\cite{LHCb:2020pxc}, where such a state might
be expected to play a significant role. The broad $\chi_{c0}(3860)$ resonance may be
contributed by several states or nonresonant backgrounds.

Then, we focus on the $\chi_{c1}(2P)$ state. This state has a strong $S$-wave coupling
to the $DD^{*}$ channel. When solving the coupled-channel
Eq.~(\ref{M=MA+Delta M}), it is seen that there are three
solutions with masses $3871$ MeV, 3888 MeV and 3990 MeV, respectively, as shown in
the upper panel of Fig.~\ref{Chic12P}.
The first solution with a mass of $3871$ MeV lies just below
the $D^0D^{*0}$ threshold. The second solution with a mass of
$3888$ MeV is just above the $D^{\pm}D^{*\mp}$ threshold.
While the third solution with a mass of 3990 MeV is heavier than
the bare state. To uncover the nature of the solutions obtained from
the coupled-channel model, we further analyze the spectral density function,
\begin{equation}\label{omega}
\begin{aligned}
\omega(M) = \frac{1}{2\pi} \frac{\Gamma}{[M - (M_{A} + \Delta M(M))]^2 + \Gamma^2/4},
\end{aligned}
\end{equation}
as adopted in the literature~\cite{Baru:2003qq,Kalashnikova:2005ui,Kalashnikova:2009gt,Wang:2023snv}.
The line shape of the spectral density function is shown in the lower panel of Fig.~\ref{Chic12P}.
It is found that the first solution with a mass of $3871$ MeV
corresponds to a very narrow state with a width of
about several MeV shown in the spectral density function. The third solution
with a mass of $3990$ MeV corresponds to a broader resonance
with a width of $\Gamma \simeq 60$ MeV. However, the second solution with
a mass of $3888$ MeV does not exhibit any resonance structures in the
spectral density function, thus, this solution is unphysical.
Similar line shape of the spectral density function was also found in the previous studies~\cite{Kalashnikova:2005ui, Giacosa:2019zxw}.
Two similar solutions are also found by the other coupled-channel
analysis of the $\chi_{c1}(2P)$~\cite{Zhou:2017dwj,Ortega:2009hj,Kalashnikova:2005ui, Giacosa:2019zxw}.
Considering the integral intervals as shown in the lower panel of Fig.~\ref{Chic12P},
from the spectral density function, one can estimate the $c\bar{c}$ core
components for the resonance structures. Our results are given in Table~\ref{Chic12Pcomponents}.
The narrow state with a mass of $3871$ MeV is dominated by the $DD^{*}$ component,
while the $c\bar{c}$ core component is only $\sim 10\%$.
The broad resonance with a mass of $3990$ MeV is a $c\bar{c}$ dominant state,
the $c\bar{c}$ core component is estimated to be $~60\%$.
The narrow state favors the famous $\chi_{c1}(3872)$ resonance.
The nature of $\chi_{c1}(3872)$ originating from the $\chi_{c1}(2P)$ is also suggested in the literature~\cite{Zhou:2017dwj,Zhou:2017txt,Duan:2020tsx,Pennington:2007xr,Li:2009ad,Meng:2007cx,Ferretti:2014xqa,
Ferretti:2013faa,Meng:2014ota,Coito:2012vf,Wang:2023ovj}.
To confirm the nature of $\chi_{c1}(3872)$, it is crucial to look for
the broad state with a mass of about $3990$ MeV in the $DD^*$ final state.

Finally, we focus on the $\chi_{c2}(2P)$ state. The unquenched
coupled-channel effects play a significant role as well.
The $D^*D^*$ and $DD^*$-loops lower the bare
mass of the $\chi_{c2}(2P)$, 4002 MeV, to the physical point $\sim 3947$ MeV.
The mass shift reaches up to a fairly large value $\Delta M\simeq -55$ MeV.
Considering the $\chi_{c2}(3930)$ resonance as $\chi_{c2}(2P)$, the measured mass is well
described within the unquenched framework, which can be clearly seen from Fig.~\ref{spectrum}.
Within the unquenched picture, the mass gap between $\chi_{c2}(2P)$ and
$\chi_{c0}(2P)$ is predicted to be $\sim 40$ MeV, which is much smaller than
$\sim60-120$ MeV predicted in the quenched quark model (see Table~\ref{ccbar mass spectrum}).
Our predictions are consistent with that in Ref.~\cite{Duan:2020tsx}.
Taking $\chi_{c2}(3930)$ as the $\chi_{c2}(2P)$ state,
one find the decay width $\Gamma\simeq 21$ MeV predicted in theory is also compatible with
the data $\Gamma_{exp}=35.2\pm 2.2$ MeV~\cite{ParticleDataGroup:2022pth}.
The $\chi_{c2}(2P)$ state dominantly decays into both $DD$ and $DD^*$ channels,
the partial width ratio between them is predicted to be
\begin{eqnarray}
\frac{\Gamma(DD^*)}{\Gamma(DD)}\simeq 0.74.
\end{eqnarray}
The $DD$ decay mode has been observed by the Belle and \emph{BABAR} experiments~\cite{ParticleDataGroup:2022pth}.

In summary, the unquenched coupled-channel effects are crucial
for uncovering the puzzles in the $\chi_{cJ}(2P)$ states.
Two resonance structures are found when considering
coupled-channel effects for the bare $\chi_{c1}(2P)$ state.
The $\chi_{c1}(3872)$ may correspond to the low-mass resonance dominated
by the $DD^*$ component.
The $\chi_{c0}(3930)$ and $\chi_{c2}(3930)$ resonances can
be well explained with the assignments $\chi_{c0}(2P)$ and $\chi_{c2}(2P)$, respectively, when the unquenched coupled-channel
effects are properly included.
The broad structure $\chi_{c0}(3860)$ cannot be explained as $\chi_{c0}(2P)$,
which may be contributed by several states or nonresonant backgrounds.

\subsubsection{$3P$-wave states}

For the $3P$-wave states, from Table~\ref{ccbar mass spectrum} one can see that
the mass corrections due to the coupled-channel effects are not significant.
The magnitude of the mass shifts is estimated to be within 30 MeV.
Except the $\chi_{c0}(3P)$ has a relatively low
mass $4234$ MeV, the masses for the other three states
$\chi_{c1,2}(3P)$ and $h_{c}(3P)$ are predicted to be around $\sim 4.3$ GeV.
Our results are consistent with the those predicted with linear potentials
~\cite{Barnes:2005pb,Deng:2016stx,Ebert:2011jc}, however, are
about $100$ MeV larger than those predicted with screened
potentials~\cite{Wang:2019mhs,Li:2009zu}.

The strong decay properties of
the $3P$-wave states are also studied, the results have been given in Table~\ref{2P3P}.
The $\chi_{c1}(3P)$ and $h_{c}(3P)$ are predicted to be narrow states with a comparable width of
$\sim 30$ MeV. While the $\chi_{c0}(3P)$ and $\chi_{c2}(3P)$ are predicted to
be moderate width states with a comparable width of $\sim 60$ MeV.
The decay properties are roughly consistent with the predictions with a linear potential model in previous work of
our group~\cite{Gui:2018rvv}. The $\chi_{c0}(3P)$ may have good potentials to be observed in
the $D^*D^*$ and $D_sD_s$ channels, the branching fractions for these two channels are predicted
to be $68\%$ and $12\%$, respectively. The $\chi_{c2}(3P)$ may have a potential to be observed in
the $DD^*$ and $D_s^*D_s^*$ channels, the predicted branching fractions may reach up to $\sim10\%$.
The $\chi_{c1}(3P)$ may have a good potential to be observed in
the $D_sD_s^*$ and $D^*_sD^*_s$ channels, the branching fractions may reach up to $\sim20\%$.
The $h_{c}(3P)$ mainly decays into the $DD^*$, $D_sD^*_s$, $D^*D^*$, and $D^*_sD^*_s$
channels with a comparable branching fraction $\sim20-30\%$.

The $\chi_{c1}(4274)$ (known as $X(4274)$) observed in
the $J/\psi\phi$ channel in the decay $B^+\to J/\psi \phi K^+$
at CDF~\cite{CDF:2011pep} and LHCb~\cite{LHCb:2016axx,LHCb:2016nsl,LHCb:2021uow} may be a good candidate
of $\chi_{c1}(3P)$. The averaged mass and width of $\chi_{c1}(4274)$, $M_{exp}=4286^{+8}_{-9}$ MeV and
$\Gamma_{exp}=51\pm 7$ MeV~\cite{ParticleDataGroup:2022pth}, are consistent with the our predictions,
$M=4287$ MeV and $\Gamma\simeq 34$ MeV. The $\chi_{c1}(4274)$ is
also suggested to be assigned as $\chi_{c1}(3P)$ in Refs.~\cite{Lu:2016cwr,Duan:2021alw,Gui:2018rvv,Wang:2022dfd,Ferretti:2020civ}.
The $\chi_{c1}(4274)$, as the $\chi_{c1}(3P)$ assignment, should be observed
in its main decay channels $D_sD_s^*$ and $D^*_sD^*_s$ via the $B^+\to D^{(*)}_sD^{(*)}_s K^+$ decays.

The $X(4350)$ structure observed in the $\gamma\gamma \to J/\psi\phi$ process at Belle~\cite{Belle:2009rkh} may be a good candidate
of $\chi_{c2}(3P)$. The measured mass and width are $M_{exp}=4350.6^{+4.6}_{-5.1}\pm 0.8$ MeV and
$\Gamma_{exp}=13^{+18}_{-9}\pm 4$ MeV, respectively. Assigning $X(4350)$ to $\chi_{c2}(3P)$, the predicted mass, $M\simeq 4337$ MeV
is in good agreement with the data, while the predicted width $\Gamma \simeq 56$ MeV
is slightly broader than the data. In Refs.~\cite{Gui:2018rvv,Liu:2009fe,Wang:2022dfd}, such a possible assignment
was also considered. To confirm this assignment future experimental search for
its decays into $DD^*$ and $D_s^*D_s^*$ are strongly recommended.

Finally, it should be mentioned that recently the LHCb collaboration carried
out observations of the $B^+\to D^{+}_sD^{-}_s K^+$ decay~\cite{LHCb:2022dvn,LHCb:2022aki},
there seems to be a bump structure around $4.3$ GeV in the $D^{+}_sD^{-}_s$
invariant-mass spectrum, which may be contributed by the $\chi_{c0}(3P)$ state. With more statistics,
this state is most likely to be established in the $D_sD_s$ channel by using the
$B^+\to D^{+}_sD^{-}_s K^+$ decay in forthcoming experiments.

As a whole, it's time to establish the $3P$-wave states, which have a relatively
narrow width of $\sim 20-60$ MeV.
The $\chi_{c1}(4274)$ and $X(4350)$ are good candidates of
the $\chi_{c1}(3P)$ and $\chi_{c2}(3P)$, respectively.
Some weak signals of $\chi_{c0}(3P)$ may have been found in the recent LHCb experiments.
The $D^{(*)}_sD^{(*)}_s$ may be good channel to establish the $3P$-wave states.

\subsubsection{$4P$-wave states}

For the $4P$-wave states, from Table~\ref{ccbar mass spectrum} one can see that
the mass corrections due to the coupled-channel effects are significant.
The magnitude of the mass shifts is estimated to be $\sim 50-90$ MeV.
Except the $\chi_{c0}(4P)$ has a relatively low
mass of $\sim 4.5$ GeV, the masses for the other three states
$\chi_{c1,2}(4P)$ and $h_{c}(4P)$ are predicted to be around $4.6$ GeV.
There are some studies of the higher $4P$-wave states within some quenched~\cite{Gui:2018rvv,Sultan:2014oua,Cao:2012du,Ebert:2011jc,Soni:2017wvy,Chaturvedi:2019usm,Mansour:2021rru,Fang:2022bft} and
unquenched~\cite{Duan:2021alw,Ferretti:2021xjl} quark models.
Our results are close to the predictions with a linear potential
in Refs.~\cite{Gui:2018rvv,Sultan:2014oua,Cao:2012du}, however, is
about $100-200$ MeV smaller than the predictions in
Refs.~\cite{Ebert:2011jc,Soni:2017wvy,Chaturvedi:2019usm},
and about $200$ MeV higher than the predictions with screened
potentials~\cite{Wang:2019mhs} and other modified confinement
potentials~\cite{Mansour:2021rru,Fang:2022bft}.

The strong decay properties of the $4P$-wave states are also studied,
the results have been given in Table~\ref{4P}.
The $\chi_{c0}(4P)$ and $\chi_{c2}(4P)$ are predicted to
be moderate width states with a comparable width of $\sim 70-90$ MeV.
While the $\chi_{c1}(4P)$ and $h_{c}(4P)$ have a slightly broader width of
$\sim 100$ MeV. The decay properties are roughly consistent with the predictions with a linear potential model
in the previous work of our group~\cite{Gui:2018rvv}. The $4P$-wave states mainly decay into the
$1P$-wave and/or $2S$-wave $D$-meson excitations by emitting a
$D$ or $D^*$ meson. The rates decaying into the OZI-allowed
$D^{(*)}D^{(*)}$, $D^{(*)}_sD^{(*)}_s$ channels are often small.

The charmonium-like resonance $\chi_{c0}(4500)$ (known as $X(4500)$) listed in RPP is a good
candidate of $\chi_{c0}(4P)$. This resonance is found by LHCb in the $J/\psi \phi$
final state via the $B^+\to J/\psi \phi K^+$ decay~\cite{LHCb:2016axx,LHCb:2021uow,LHCb:2016nsl}.
The newly measured mass and width
of $\chi_{c0}(4500)$ are $M_{exp}=4474\pm 6$ MeV and $\Gamma_{exp}=77\pm6^{+10}_{-8}$ MeV,
respectively~\cite{LHCb:2021uow}, which are consistent with theoretical predictions $M\simeq4486$ MeV and $\Gamma\simeq 106$ MeV.
If $\chi_{c0}(4500)$ corresponds to $\chi_{c0}(4P)$ indeed, the decay
rate into $D_s^+D_s^-$ channel is sizeable, the branching fraction
is estimated to be $\sim 1\%$. Thus, the $\chi_{c0}(4500)$ resonance
should be established in the $D_s^+D_s^-$ final state by using the
$B^+\to D_s^+D_s^- K^+$ decay. Recently, this process has been observed by LHCb~\cite{LHCb:2022dvn,LHCb:2022aki}.
There seems to be a vague bump structure around $4.5$ GeV in
the $D^{+}_sD^{-}_s$ invariant mass spectrum. With more statistics,
the $\chi_{c0}(4500)$ is most likely to be established in $B^+\to D_s^+D_s^- K^+$.

\subsubsection{$5P$-wave states}

For the higher $5P$-wave states, from Table~\ref{ccbar mass spectrum} one can see that
the mass corrections due to the coupled-channel effects are significant.
The magnitude of the mass shifts is estimated to be $\sim 30-60$ MeV.
Except the $\chi_{c0}(5P)$ has a relatively low
mass $4780$ MeV, the masses for the other three states
$\chi_{c1,2}(5P)$ and $h_{c}(5P)$ are predicted to be around $4.9$ GeV.
There are some studies of the higher $5P$-wave states within some quenched
~\cite{Gui:2018rvv,Sultan:2014oua,Soni:2017wvy,Mansour:2021rru} and
unquenched~\cite{Duan:2021alw,Ferretti:2021xjl} quark models.
Strong model dependencies exist in the predictions.
Our results are close to those predicted with a linear potential
in Refs.~\cite{Ferretti:2021xjl,Gui:2018rvv,Sultan:2014oua}, however, is
about $100-200$ MeV smaller than those predicted in
Refs.~\cite{Ebert:2011jc,Soni:2017wvy},
and about $300-400$ MeV higher than the those predicted with
a unquenched quark model~\cite{Duan:2021alw} and other modified confinement
potentials~\cite{Mansour:2021rru}.

The strong decay properties of the $5P$-wave states are also studied,
the results have been given in Tables~\ref{53P053P2} and~\ref{51P153P1}.
It is found that the $\chi_{c0}(5P)$ has a width of $\Gamma\simeq 70$ MeV,
while the other three states $\chi_{c1,2}(5P)$ and $h_{c}(5P)$ have a slightly
broader width of $\sim 100$ MeV. The strong decay properties
are roughly consistent with those predicted
with a linear potential model in the previous work of
our group~\cite{Gui:2018rvv}. The $5P$-wave states mainly decay into the
$1P$/$2P$-wave and/or $2S$-wave $D$-meson excitations by emitting a
light $D$ or $D^*$ meson. The rates decaying into the OZI-allowed
$D^{(*)}D^{(*)}$, $D^{(*)}_sD^{(*)}_s$ channels are negligibly small.

The charmonium-like resonance $\chi_{c0}(4700)$ (known as $X(4700)$)
listed in RPP may be a good candidate of $\chi_{c0}(5P)$.
The $\chi_{c0}(4700)$ was first observed by LHCb in the $J/\psi \phi$
invariant mass spectrum via the $B^+\to J/\psi \phi K^+$ decays
in 2016~\cite{LHCb:2016nsl,LHCb:2016axx}, and confirmed in the
same process with more statistics in 2021~\cite{LHCb:2021uow}.
The lately measured mass and width are
$M_{exp}=4694\pm 6^{+16}_{-3}$ MeV and $\Gamma_{exp}=87\pm 8\pm^{+16}_{-6}$ MeV,
respectively~\cite{LHCb:2021uow}. By using the $B^0_s\to J/\psi \phi \pi^+\pi^-$ decays,
the LHCb collaboration also observed a similar structure in the
$J/\psi \phi$ channel with a mass of $M_{exp}=4741\pm 6 \pm 6$ MeV and
$\Gamma_{exp}=53\pm 15\pm 11$ MeV, respectively~\cite{LHCb:2020coc}. The measured mass and
width of $\chi_{c0}(4700)$ are consistent with our predictions,
$M\simeq 4780$ MeV and $\Gamma \simeq 71$ MeV, with
the $\chi_{c0}(5P)$ assignment. The coupled-channel effects
are crucial for understanding the mass of $\chi_{c0}(4700)$.
These unquenched effects can lower the bare mass of $\chi_{c0}(5P)$
with a value of $\sim60$ MeV. If $\chi_{c0}(4700)$ corresponds
to $\chi_{c0}(5P)$ indeed, it is most likely to be
observed in the $DD_1(2420)$ channel via the
$B^+\to \chi_{c0}(4700) K^+$ process. The branching
fraction for $\chi_{c0}(4700)\to DD_1(2420)$ may
reach up to $\sim7\%$.


\subsection{$D$-wave states}\label{D-wave}

The mass spectrum up to $5D$-wave states is given in Table~\ref{ccbar mass spectrum} and
also shown in Fig.~\ref{spectrum}. The strong decay properties for the
$D$-wave states are given in Tables~\ref{1D2D}-\ref{51D253D2}.
The mass shifts due to the unquenched coupled-channel effects on the are predicted to be in the range
of $\sim(-70,-20)$ MeV. The high $D$-wave states have a relatively narrow
width of $\sim 10s-100$ MeV, although many decay channels are fully opened.

\subsubsection{$1D$-wave states}\label{1D-wave}

For the low-lying $1D$ states, except the spin singlet $\eta_{c2}(1D)$,
all the spin triplets, $\psi(1D)$, $\psi_2(1D)$, and $\psi_3(1D)$, have been well established.
Assigning the $\psi(3770)$, $\psi_2(3823)$ and $\psi_3(3842)$ as the spin triplets of
the $1D$-wave states, from Fig.~\ref{spectrum} it is seen that their masses are consistent
with the quark model expectations.

The decays of both $\psi(3770)$ and $\psi_3(3842)$ are
governed by the $DD$ channel, their decay widths are consistent with the quark model
expectations as well (see Table~\ref{1D2D}). For $\psi_2(3823)$, the OZI-allowed
two-body strong decays are kinematic forbidden, thus, its decays is dominated
by the electromagnetic transitions. The measured electromagnetic decay properties
are also consistent with the theoretical predictions~\cite{Deng:2016stx,BESIII:2021qmo}.

How to find the missing spin singlet $\eta_{c2}(1D)$ is still a
challenge in experiments. It may be produced via the
$B\to \eta_{c2}(1D) K$ decay as suggested in Refs.~\cite{Eichten:2002qv,Xu:2016kbn,Fan:2009cj},
and established by the the two-photon cascade decay process
$\eta_{c2}(1D)\to h_c(1P)\gamma \to \eta_c \gamma\gamma$~\cite{Deng:2016stx}.

\subsubsection{$2D$-wave states}\label{2D-wave}

In the $2D$-wave states, only the $\psi(2D)$ with $J^{PC}=1^{--}$
has been established. The $\psi(4160)$ resonance is usually assigned as
the $\psi(2D)$ state. With this assignment, the measured
width $\Gamma_{exp}=70\pm 10$ MeV are consistent with
the theoretical prediction $\Gamma\simeq 62$ MeV.
However, the mass $M_{exp}=4191\pm 5$ MeV extracted from experiments
is about $40$ MeV higher than most of the predictions in theory (See Table~\ref{ccbar mass spectrum}).
This state dominantly decays into $D^*D^*$ channel with a branching
fraction of $\sim 70\%$, while it also has sizeable decay rates into
$DD$ and $D_sD_s^*$ channels with a comparable branching
fraction of $\sim 10\%$ (see Table~\ref{1D2D}). The branching fraction ratios
predicted in the present work and other works~\cite{Gui:2018rvv,Eichten:2005ga,
Segovia:2013kg} are very different from the old
observations at \emph{BABAR}~\cite{BaBar:2009elc}.
In these $P$-wave decay channels there exist obvious interfering effects
between $\psi(4040)$ and $\psi(4160)$~\cite{Gui:2018rvv}. A coherent partial wave
analysis combining all these exclusive channels is suggested to be carried out for
extracting the resonance parameters and branching fractions for these two states.

The other three states $\psi_3(2D)$, $\eta_{c2}(2D)$ and $\psi_2(2D)$ have a
similar mass of $\sim 4.2$ GeV. The corrections
to the bare masses due to the coupled-channel effects
are not significant, the magnitude of the mass shifts is
within $20$ MeV. Our predicted masses for these $2D$-wave
states are in agreement with most of the predictions in the
literature, such as Refs.~\cite{Ebert:2011jc,Godfrey:1985xj,Deng:2016stx,Barnes:2005pb}.
The theoretical strong decay properties have been given in
Table~\ref{1D2D}. It is found that $\psi_3(2D)$, as the narrowest state in the $2D$-wave
states, has a width of $\Gamma\simeq 38$ MeV, while mainly decays into
the $DD^*$ and $D^*D^*$ channels with branching fractions $\sim34\%$ and $\sim47\%$, respectively.
Both the $J^P=2^-$ states, $\psi_2(2D)$ and $\eta_{c2}(2D)$, have a comparable width of
$\Gamma\simeq 60$ MeV. The $\psi_2(2D)$ ($\eta_{c2}(2D)$) mainly decays into
the $DD^*$, $D_sD^*_s$, and $D^*D^*$ channels with branching fractions
$\sim30\%$ ($\sim35\%$), $\sim17\%$ ($\sim11\%$) and $\sim52\%$ ($\sim43\%$), respectively.

Some signals of $\eta_{c2}(2D)$ may have been observed in experiments. In 2021,
the LHCb collaboration observed a new resonance $X(4160)$ with a significance of $4.8$ $\sigma$
in the $J/\psi \phi$ final state by using an amplitude analysis of
the $B^+\to J/\psi \phi K^+$ decays~\cite{LHCb:2021uow}. The $J^{PC}=2^{-+}$ assignment is favored
over other assignments with a significance of more than $4.8$ $\sigma$.
The observed mass and width are $M_{exp}=4146\pm 18\pm33$ MeV and
$\Gamma_{exp}=135\pm 28^{+25}_{-30}$ MeV, respectively, which are consistent
with the state observed in $e^+e^-\to J/\psi X$ with $X\to D^*D^*$ by Belle~\cite{Belle:2007woe}.
Considering $X(4160)$ as the $J^{PC}=2^{-+}$ state $\eta_{c2}(1D)$,
both the observed mass and width are consistent with the
theoretical predictions, $M\simeq 4192$ MeV and $\Gamma\simeq 60$ MeV.
To confirm the nature of $X(4160)$, the other two dominant decay modes,
$DD^*$ and $D_sD^*_s$, are expected to be searched for in future experiments.

\subsubsection{$3D$-wave states}\label{3D-wave}

The $3D$-wave states, $\psi(3D)$, $\psi_3(3D)$, $\eta_{c2}(3D)$ and $\psi_2(3D)$,
may highly degenerate with each other in a very narrow mass range of $\sim4460-4470$ MeV.
From Table~\ref{ccbar mass spectrum}, one can see that the
coupled-channel effects have a significant correction to the bare masses of the
$3D$-wave states with $J^{P}=2^-$ and $3^-$.
The magnitude of the mass shifts due to these unquenched effects can
reach up to about $60-70$ MeV. The masses of the $3D$-wave states predicted in the
present work are consistent with the those predicted with linear potentials
~\cite{Deng:2016stx,Sultan:2014oua,Kher:2018wtv} and a Martin-like potential~\cite{Shah:2012js}, however, are
about $130$ MeV larger than those predicted with screened
potentials~\cite{Deng:2016stx,Wang:2019mhs,Li:2009zu}, while about
$100$ MeV smaller than those predicted with other potential models~\cite{Ebert:2011jc,Soni:2017wvy}.

The strong decay properties of the $3D$-wave states are given in Table~\ref{3D}.
From the table, one can see that the widths for the $3D$-wave states
are not broad. In these states, the $\psi(3D)$ is the broadest state
with a width of $\Gamma\simeq 119$ MeV. This state dominantly decays into
$DD_0(2550)$, $DD_2^*(2460)$, $DD_1(2420)$, $D^*D_1(2420)$ channels with
branching fractions $\sim 24\%$, $\sim 9\%$, $\sim 27\%$, and $\sim 18\%$, respectively.
The $\psi_3(3D)$ has a narrow width of $\Gamma\simeq 38$ MeV, and mainly
decays into $DD_2^*(2460)$, $DD_1(2420)$, $D^*D_1(2430)$ channels with
branching fractions $\sim 17\%$, $\sim 15\%$, and $\sim 24\%$, respectively.
The $\psi_2(3D)$ has a moderate width of $\Gamma\simeq 65$ MeV, and mainly
decays into $DD_2^*(2460)$ and $D^*D_1(2420)$ channels with
a comparable branching fraction of $\sim 30\%$. The $\eta_{c2}(3D)$ is
the narrowest state with a width of $\Gamma\simeq 30$ MeV. This
state has large decay rates into the $DD_2^*(2460)$,
$D^*D_1(2430)$, and $D^*D_1(2420)$ channels with branching fractions of $\sim 22\%$,
$\sim 33\%$, and $\sim 13\%$, respectively.

Recently, the BESIII collaboration observed a vector resonance
$Y(4500)$ in the line shape of the cross sections of the
$e^+e^-\to K^+K^- J/\psi$ process~\cite{BESIII:2022joj}. The mass and width are
measured to be $M_{exp}=4484.7\pm 13.3\pm24.1$ MeV and $\Gamma_{exp}=111.1\pm30.1\pm15.2$ MeV,
respectively. This state is confirmed in $e^+e^-\to D^{*0}D^{*-}\pi^+$
by BESIII~\cite{BESIII:2023cmv}.
The observed mass, width and quantum numbers
of $Y(4500)$ can be well understood in theory with
the $\psi(3D)$ assignment. If $Y(4500)$ corresponds
to $\psi(3D)$ indeed, it is most likely to be observed in the dominant decay channels, such as
$DD_2^*(2460)$, $DD_1(2420)$, $D^*D_1(2420)$, in future experiments.
The $D^{*0}D^{*-}\pi^+$ decay mode of $Y(4500)$ observed by BESIII~\cite{BESIII:2023cmv}
may be mainly contributed by $D^*D_1(2420)\to D^*D^*\pi$.

\subsubsection{$4D$-wave states}\label{4D-wave}

The $4D$-wave states are predicted to lie in the mass range of
$\sim4730-4800$ MeV (see Table~\ref{ccbar mass spectrum}).
The mass corrections due to the unquenched coupled-channel effects
are not large, the values are about $30$ MeV. The masses of the $4D$-wave states predicted in the
present work are consistent with the those predicted with linear potentials~\cite{Deng:2016stx,Sultan:2014oua}, however, are
about $130$ MeV larger than those predicted with a screened
potential~\cite{Wang:2019mhs} and a Martin-like potential~\cite{Shah:2012js}, while about
$100$ MeV smaller than those predicted with other potential models~\cite{Chaturvedi:2019usm,Ebert:2011jc,Soni:2017wvy}.

The strong decay properties of the $4D$-wave states are given in Table~\ref{4D}.
In these states, the $\psi(4D)$ and $\psi_3(4D)$ have a
relatively broader width of $\Gamma\sim 110$ MeV, while the two $J^P=2^-$ states
$\eta_{c2}(4D)$ and $\psi_2(4D)$ have a moderate width of $\Gamma\sim 80$ MeV.
The decay rates of the $4D$-wave states into the
OZI-allowed $D^{(*)}D^{(*)}$ and $D_s^{(*)}D_s^{(*)}$ channels are tiny.
The branching fractions for these channels are predicted
to be less than $1\%$. The $1^{--}$ state $\psi(4D)$ has large decay rates into the
$D^*D_1^*(2600)$, $D^*D_1(2420)$, and $DD(1D_2')$ channels with branching fractions
$\sim 37\%$, $\sim 8\%$, and $\sim 16\%$, respectively.
The $3^{--}$ state $\psi_3(4D)$ has large decay rates into the
$D^*D_1^*(2600)$, $D^*D_2^*(2460)$, and $D^*D_3^*(2750)$ channels with branching fractions
$\sim 17\%$, $\sim 12\%$, and $\sim 37\%$, respectively.
The $2^{--}$ state $\psi_2(4D)$ has large decay rates into the
$D^*D_1^*(2600)$, $D^*D_2^*(2460)$, $D^*D_1(2420)$, and $DD_3^*(2750)$ channels with branching fractions
$\sim 28\%$, $\sim 8\%$, $\sim 7\%$, and $\sim 10\%$, respectively.
The $2^{-+}$ state $\eta_{c2}(4D)$ has large decay rates into the
$D^*D_1^*(2600)$, $D^*D_2^*(2460)$, and $DD_3^*(2750)$ channels with branching fractions
$\sim 27\%$, $\sim 11\%$, and $\sim 16\%$, respectively.

Some signals of the $1^{--}$ state $\psi(4D)$ may have been seen in experiments.
Recently, a new vector resonance $Y(4710)$ was observed in $e^+e^-\to K\bar{K}J/\psi$ at BESIII~\cite{BESIII:2023wqy,BESIII:2022kcv}, the measured mass and width are $\sim 4710$ GeV and $\sim 100$ MeV, respectively.
The predicted mass and width of $\psi(4D)$, $M\simeq 4737$ MeV and $\Gamma\simeq 107$ MeV
are consistent with those of $Y(4710)$. Lately, in the Born cross sections of
$e^+e^-\to D_s^{*+}D_s^{*-}$, a structure ($Y(4790)$) with a mass
of $\sim 4.7-4.8$ GeV and a width of $\sim 27-60$ MeV was observed by BESIII~\cite{BESIII:2023wsc}. This might be
the same state observed in $e^+e^-\to K\bar{K}J/\psi$.
It should be mentioned that the $\psi(4D)$ may highly overlap
with $\psi(5S)$ in the mass range of $\sim 4.7$ GeV.
Only a small mass gap ($\sim 55$ MeV) between them is predicted
in theory. Both $\psi(4D)$ and $\psi(5S)$ have a comparable
decay rate, $\mathcal{O}(10^{-3})$, into the $D_s^{*+}D_s^{*-}$
channel. Thus, the structure observed at around
$4.7-4.8$ GeV may be contributed by $\psi(4D)$ and/or $\psi(5S)$.
Further observations with more data samples may be useful to
uncover the nature of the $Y(4710)$ structure.

\subsubsection{$5D$-wave states}\label{5D-wave}

The $5D$-wave states are predicted to lie in the mass range of
$\sim4988-5060$ MeV (see Table~\ref{ccbar mass spectrum}).
The mass corrections due to the unquenched coupled-channel effects
are about $30-40$ MeV. The masses of the $5D$-wave states predicted in the
present work are consistent with the those predicted within a linear potential model
~\cite{Sultan:2014oua}.

The strong decay properties of the $5D$-wave states are given in
Tables~\ref{53D153D3} and~\ref{51D253D2}. It is found that
these high $5D$-wave states have a relatively narrow width
of $\sim30-60$ MeV. The decay rates of the $5D$-wave states into the
OZI-allowed $D^{(*)}D^{(*)}$ and $D_s^{(*)}D_s^{(*)}$ channels are tiny.
The $1^{--}$ state $\psi(5D)$ has relatively large decay rates into the
$D^*D_1^*(2600)$ and $D_1(2420)D_1(2420)$ channels with branching fractions
$\sim 6\%$ and $\sim 8\%$, respectively.
The $3^{--}$ state $\psi_3(5D)$ has relatively large decay rates into the
$D^*D_3^*(2750)$ and $D_2^*(2460)D_2^*(2460)$ channels with branching fractions
$\sim 14\%$ and $\sim 9\%$, respectively. The $2^{--}$ state $\psi_2(5D)$ has relatively large decay rates into the
$D_{s0}^*(2317)D_{s1}(2460)$ and $D_2^*(2460)D_2^*(2460)$ channels with branching fractions
$\sim 17\%$ and $\sim 4\%$, respectively. The $2^{-+}$ state $\eta_{c2}(5D)$ has relatively large decay rates into the
$D_2^*(2460)D_2^*(2460)$ and $D_2^*(2460)D_1(2420)$ channels with branching fractions
$\sim 7\%$ and $\sim 5\%$, respectively.

\section{Summary}\label{SUM}

In this work, the mass spectrum and strong decay properties of the $S$-, $P$-, and $D$-wave
charmonium states up to mass region of $\sim 5.0$ GeV are
systematically studied within the unquenched quark model including
coupled-channel effects from all of the OZI-allowed opened charmed meson channels.
We can obtain a good description of both the masses and widths
for the well-established states in the charmonium spectrum.
Although many decay channels are fully opened for the
higher charmonium states, they are relatively narrow states.
Their widths scatter in the range of $\sim 10s-100$ MeV.
We expect our study can provide a useful reference for
establishing an abundant charmonium spectrum.
Some key results from this study are emphasized as follows.
\begin{itemize}
\item The magnitude of mass shifts of the bare states due to the coupled-channel
effects are estimated to be within $10s$ MeV. The
mass shifts do not show an obvious increasing trend from lower
levels to higher ones as that predicted in the screened potential models.

\item Two resonance structures are found when considering
coupled-channel effects for the bare $\chi_{c1}(2P)$ state.
The $\chi_{c1}(3872)$ favors the low-mass resonance dominated
by the $DD^*$ component.

\item The $\chi_{c0}(3915)$ resonance can be well understood
with the dressed $\chi_{c0}(2^3P_0)$ states in the coupled-channel model.

\item The vector resonances $\psi(4415)$ and $\psi(4660)$ favor
the $\psi(4S)$ and $\psi(5S)$ assignments, respectively.

\item The newly observed vector states $Y(4500)$ and $Y(4710)$
at BESIII may favor the $\psi(3D)$ and $\psi(4D)$ assignments, respectively.
However, $Y(4710)$ as the $\psi(5S)$ assignment cannot be excluded.

\item The $\chi_{c0}(4500)$ and $\chi_{c0}(4700)$ resonances observed by LHCb
favor the $\chi_{c0}(4P)$ and $\chi_{c0}(5P)$ assignments, respectively.

\item The $\chi_{c1}(4274)$ resonance observed at CDF and LHCb
and $X(4350)$ observed at Belle may be good candidates of $\chi_{c1}(3P)$
and $\chi_{c2}(3P)$, respectively.

\item The $X(4160)$ resonance observed at LHCb and Belle favor
the $\eta_{c2}(1D)$ assignment

\item  The $\chi_{c0}(3860)$, $\psi(4230)$, $Y(4360)$, $\chi_{c1}(4140)$,
and $\chi_{c1}(4685)$ resonances listed in RPP cannot be accommodated by the charmonium spectrum.
\end{itemize}

\begin{table*}[htbp]
\begin{center}
\caption{The mass shifts and strong decay widths for the $5^1S_0$ and $6^1S_0$ charmonium states (in MeV).
The ``\ding{55}'' is labeled the strong decay channels which are
not open or forbidden. These channels are considered to have no contributions to
the mass shifts and decay widths. The ``$...$'' stands for the predicted
values are negligibly small.}
			\label{5S}
			\tabcolsep=0.10cm
			\renewcommand\arraystretch{0.5}
			\scalebox{1.0}{
\begin{tabular}{cccccccccccccccccc}
					\bottomrule[1.0pt]\bottomrule[1.0pt]
					&\multicolumn{2}{c}{$5^1S_0$}  &  &\multicolumn{2}{c}{$6^1S_0$}&
					&\multicolumn{2}{c}{$5^1S_0$}  &  &\multicolumn{2}{c}{$6^1S_0$}
					\\
\cline{2-3}\cline{5-6}\cline{8-9}\cline{11-12}
	Channel				&$\Delta M$~[4708]  &$\Gamma$~[4662]  &   &$\Delta M$~[4983]  &$\Gamma$~[4902] &
	\emph{Continue}		&$\Delta M$~[4708]  &$\Gamma$~[4662]  &   &$\Delta M$~[4983]  &$\Gamma$~[4902]     \\
					\bottomrule[0.4pt]
					$DD^*$                            &$-0.06$     &$...$    &      &$-0.03$       &$...$
					&$D_s^*D_{s1}(2536)$              &$-2.90$     &$0.52$   &      &$-0.01$       &$0.29$\\
					$D_sD_s^*$                        &$-0.01$     &$0.27$   &      &$-0.01$       &$0.03$
					&$D_1^*(2600)D_1(2430)$           &\ding{55}   &\ding{55}&      &$-3.00$       &\ding{55}\\
					$D^*D^*$                          &$+0.53$     &$1.12$   &      &$+0.09$       &$0.20$
					&$D_1^*(2600)D_1(2420)$           &\ding{55}   &\ding{55}&      &$-3.86$       &\ding{55}\\
					$D_s^*D_s^*$                      &$-0.23$     &$0.36$   &      &$-0.03$       &$0.06$
					&$DD(1^3D_1)$                     &$-0.29$     &$2.85$   &      &$-0.50$       &$0.48$\\
					$D_0(2550)D^*$                    &$-4.37$     &$5.12$   &      &$-0.09$       &$2.93$
					&$D_sD_s(1^3D_1)$                 &\ding{55}   &\ding{55}&      &$-0.01$       &$...$\\
					$D(3^1S_0)D^*$                    &\ding{55}   &\ding{55}&      &$-2.38$       &\ding{55}
					&$DD(2^3D_1)$                     &\ding{55}   &\ding{55}&      &$-0.80$       &\ding{55}\\
\end{tabular}}
		\end{center}
	\end{table*}

	\begin{table*}[htbp]
		\begin{center}
			\caption{The mass shifts and strong decay widths for the $5^3S_1$ and $6^3S_1$ charmonium states (in MeV).
The ``\ding{55}'' is labeled the strong decay channels which are
not open or forbidden. These channels are considered to have no contributions to
the mass shifts and decay widths. The ``$...$'' stands for the predicted
values are negligibly small.}
			\label{6S}
			\tabcolsep=0.10cm
			\renewcommand\arraystretch{0.50}
			\scalebox{1.0}{
				\begin{tabular}{cccccccccccccccccc}
					\bottomrule[1.0pt]\bottomrule[1.0pt]
					&\multicolumn{2}{c}{$5^3S_1$}  &  &\multicolumn{2}{c}{$6^3S_1$}&
					&\multicolumn{2}{c}{$5^3S_1$}  &  &\multicolumn{2}{c}{$6^3S_1$}
					\\
					 \cline{2-3}\cline{5-6}\cline{8-9}\cline{11-12}
					Channel				&$\Delta M$~[4733]  &$\Gamma$~[4682]  &   &$\Delta M$~[5005]  &$\Gamma$~[4941] &
	\emph{Continue}		&$\Delta M$~[4733]  &$\Gamma$~[4682]  &   &$\Delta M$~[5005]  &$\Gamma$~[4941]     \\
					\bottomrule[0.4pt]
					$DD$                             &$-0.07$   &$0.17$ &      &$-0.02$ &$0.06$                   &$D^*D(2P'_1)$                   &\ding{55} &\ding{55} &      &$-6.71$ &$1.37$   \\
					$D_sD_s$                         &$-0.03$   &$0.01$ &      &$-0.01$ &$0.01$                   &$D^*D_1(2420)$                  &$-2.40$   &$6.28$ &         &$-0.01$ &$0.99$  \\
					$DD^*$                           &$-0.20$   &$0.08$ &      &$-0.07$ &$0.10$                   &$D^*(2600)D_1(2420)$            &\ding{55} &\ding{55}&       &$-2.36$ &\ding{55} \\                                 			
					$D_sD_s^*$                       &$...$     &$0.10$ &      &$-0.01$ &$...$                    &$DD(1^3D_1)$                    &$+0.01$   &$0.21$ &         &$-0.05$ &$0.22$   \\                                  			
					$D^*D^*$                         &$+0.45$   &$0.56$ &      &$+0.05$ &$...$                    &$DD(2^3D_1)$                    &\ding{55} &\ding{55} &      &$-0.15$ &\ding{55}  \\                                			
					$D_s^*D_s^*$                     &$-0.20$   &$0.50$ &      &$-0.02$ &$0.08$                   &$DD(1^3D_3)$                    &$-5.19$   &$1.94$ &         &$-0.14$ &$0.53$  \\
\end{tabular}}
		\end{center}
	\end{table*}

	\begin{table*}
		\begin{center}
			\caption{The mass shifts and strong decay widths for the  $2P$- and $3P$-wave charmonium states (in MeV).
The ``\ding{55}'' is labeled the strong decay channels which are
not open or forbidden. These channels are considered to have no contributions to
the mass shifts and decay widths. The ``$...$'' stands for the predicted
values are negligibly small. }
			\label{2P3P}
			\tabcolsep=0.16cm
			\renewcommand\arraystretch{0.50}
			\scalebox{1.0}{
				\begin{tabular}{cccccccccccccccccc}
					\bottomrule[1.0pt]\bottomrule[1.0pt]
					&&\multicolumn{2}{c}{$2^3P_0$}  &  &\multicolumn{2}{c}{$2^3P_2$}  &
					&\multicolumn{2}{c}{$2^1P_1$}   &  &\multicolumn{2}{c}{$2^3P_1$} \\
					 \cline{3-4}\cline{6-7}\cline{9-10}\cline{12-13}
					&  &$\Delta M$   &$\Gamma$     &  &$\Delta M$   &$\Gamma$
					&  &$\Delta M$   &$\Gamma$     &  &$\Delta M$   &$\Gamma$     \\
					Channel       &  &$[3865]$     &$as~\chi_{c0}(3915)$    &  &$[4002]$    &$as~\chi_{c2}(3930)$
					&  &$[3968]$     &$[3961]$                &  &$[3965]$    &$as~\chi_{c1}(3872)/[3990]$   \\
					\bottomrule[0.4pt]
					$DD$                       &  &$+39.86$            &$15.82$                         &  &$-0.23$       &$12.10$                      &   &\ding{55}   &\ding{55}               &   &\ding{55}        &\ding{55}              \\
					$D_sD_s$                   &  &\ding{55}          &\ding{55}                        &  &$-1.24$       &\ding{55}                    &   &\ding{55}   &\ding{55}               &   &\ding{55}        &\ding{55}              \\
					$DD^*$                     &  &\ding{55}          &\ding{55}                        &  &$-19.57$      &$8.86$                       &   &$-6.82$     &$66.15$                 &   &$-93.92/+24.79$   &\ding{55}$\bm{/59.56}$  \\
					$D_sD_s^*$                 &  &\ding{55}          &\ding{55}                        &  &\ding{55}     &\ding{55}                    &   &\ding{55}   &\ding{55}               &   &\ding{55}        &\ding{55}               \\
					$D^*D^*$                   &  &\ding{55}          &\ding{55}                        &  &$-33.60$      &\ding{55}                    &   &\ding{55}   &\ding{55}               &   &\ding{55}        &\ding{55}               \\
\end{tabular}}
		\end{center}
	\end{table*}
	
	\begin{table*}
		\begin{center}
			\caption{The mass shifts and strong decay widths for the $4P$-wave charmonium states (in MeV).
The ``\ding{55}'' is labeled the strong decay channels which are
not open or forbidden. These channels are considered to have no contributions to
the mass shifts and decay widths. The ``$...$'' stands for the predicted
values are negligibly small.}
			\label{4P}
			\tabcolsep=0.18cm
			\renewcommand\arraystretch{0.50}
			\scalebox{1.0}{
				\begin{tabular}{cccccccccccccccccccccccc}
					\bottomrule[1.0pt]\bottomrule[1.0pt]
					& &\multicolumn{2}{c}{$4^3P_0$}  &  &\multicolumn{2}{c}{$4^3P_2$}  &
					&\multicolumn{2}{c}{$4^1P_1$}   &   &\multicolumn{2}{c}{$4^3P_1$} \\
\cline{3-4}\cline{6-7}\cline{9-10}\cline{12-13}
					Channel			&	&$\Delta M$~[4551]  &$\Gamma$~[4489]  &   &$\Delta M$~[4660]  &$\Gamma$~[4571] &
	&$\Delta M$~[4625]  &$\Gamma$~[4573]  &   &$\Delta M$~[4625]  &$\Gamma$~[4562]     \\
					\bottomrule[0.4pt]
          $DD$                     &    &$+1.10$         &$0.34$         &  &$...$        &$0.40$         &   &\ding{55}  &\ding{55}     &   &\ding{55}    &\ding{55}    \\
					$D_sD_s$                 &    &$+0.46$         &$1.29$         &  &$-0.05$      &$0.06$         &   &\ding{55}  &\ding{55}     &   &\ding{55}    &\ding{55}     \\
					$DD^*$                   &    &\ding{55}       &\ding{55}      &  &$-0.42$      &$1.02$         &   &$-0.54$    &$1.01$        &   &$-0.02$      &$0.43$  \\
					$D_sD_s^*$               &    &\ding{55}       &\ding{55}      &  &$-0.06$      &$0.01$         &   &$+0.07$    &$0.41$        &   &$+0.23$      &$0.93$  \\
					$D^*D^*$                 &    &$+1.68$         &$7.75$         &  &$+0.26$      &$0.39$         &   &$+0.52$    &$0.76$        &   &$+0.03$      &$0.05$  \\
					$D_s^*D_s^*$             &    &$-1.06$         &$0.67$         &  &$-0.24$      &$1.25$         &   &$-0.30$    &$1.38$        &   &$-0.22$      &$1.03$  \\
\end{tabular}}
		\end{center}
	\end{table*}

	\begin{table*}[htbp]
		\begin{center}
			\caption{The mass shifts and strong decay widths for the $5^3P_0$ and $5^3P_2$ charmonium states (in MeV).
The ``\ding{55}'' is labeled the strong decay channels which are
not open or forbidden. These channels are considered to have no contributions to
the mass shifts and decay widths. The ``$...$'' stands for the predicted
values are negligibly small.}
			\label{53P053P2}
			\tabcolsep=0.10cm
			\renewcommand\arraystretch{0.50}
			\scalebox{1.0}{
				\begin{tabular}{cccccccccccccccccc}
					\bottomrule[1.0pt]\bottomrule[1.0pt]
					&\multicolumn{2}{c}{$5^3P_0$}  &  &\multicolumn{2}{c}{$5^3P_2$}&
					&\multicolumn{2}{c}{$5^3P_0$}  &  &\multicolumn{2}{c}{$5^3P_2$}
					\\
					 \cline{2-3}\cline{5-6}\cline{8-9}\cline{11-12}
					Channel				&$\Delta M$~[4839]  &$\Gamma$~[4780]  &   &$\Delta M$~[4939]  &$\Gamma$~[4882] &
	\emph{Continue}		&$\Delta M$~[4839]  &$\Gamma$~[4780]  &   &$\Delta M$~[4939]  &$\Gamma$~[4882]     \\
					\bottomrule[0.4pt]
					$DD$                           &$+0.28$     &$0.04$    &     &$+0.01$  &$0.10$                 &$D^*D(2^3P_2)$                &\ding{55}   &\ding{55}&      &$-12.12$ &\ding{55}  \\
					$D_sD_s$                       &$+0.15$     &$0.20$    &     &$-0.01$  &$0.02$                 &$D^*D_1(2430)$                &$-2.11$     &$2.79$    &      &$+0.42$  &$21.36$   \\
					$DD^*$                         &\ding{55}   &\ding{55} &     &$-0.07$  &$0.32$                 &$D^*D(2P_1)$                  &\ding{55}   &\ding{55} &      &$-3.77$  &\ding{55} \\
					$D_sD_s^*$                     &\ding{55}   &\ding{55} &     &$-0.03$  &$0.01$                 &$D_s^*D_{s1}(2460)$           &$-0.31$     &$0.13$    &      &$-0.27$  &$4.63$   \\
					$D^*D^*$                       &$+0.45$     &$1.21$    &     &$-0.11$  &$0.11$                 &$D^*D_1(2420)$                &$-1.13$     &$1.24$    &      &$+0.29$  &$9.16$   \\
					$D_s^*D_s^*$                   &$-0.20$     &$0.43$    &     &$+0.03$  &$0.17$                 &$D^*D(2P'_1)$                 &\ding{55}   &\ding{55} &      &$-3.77$  &\ding{55}  \\
\end{tabular}}
		\end{center}
	\end{table*}

	\begin{table*}[htbp]
		\begin{center}
			\caption{The mass shifts and strong decay widths for the $5^1P_1$ and $5^3P_1$ charmonium states (in MeV).
The ``\ding{55}'' is labeled the strong decay channels which are
not open or forbidden. These channels are considered to have no contributions to
the mass shifts and decay widths. The ``$...$'' stands for the predicted
values are negligibly small. }
			\label{51P153P1}
			\tabcolsep=0.10cm
			\renewcommand\arraystretch{0.50}
			\scalebox{1.0}{
				\begin{tabular}{cccccccccccccccccc}
					\bottomrule[1.0pt]\bottomrule[1.0pt]
					&\multicolumn{2}{c}{$5^1P_1$}  &  &\multicolumn{2}{c}{$5^3P_1$}&
					&\multicolumn{2}{c}{$5^1P_1$}  &  &\multicolumn{2}{c}{$5^3P_1$}
					\\
					 \cline{2-3}\cline{5-6}\cline{8-9}\cline{11-12}
					Channel				&$\Delta M$~[4905]  &$\Gamma$~[4875]  &   &$\Delta M$~[4905]  &$\Gamma$~[4856] &
	\emph{Continue}		&$\Delta M$~[4905]  &$\Gamma$~[4875]  &   &$\Delta M$~[4905]  &$\Gamma$~[4856]    \\
					\bottomrule[0.4pt]
					$DD^*$                        &$-0.15$     &$0.37$    &     &$-0.05$  &$0.16$                  &$D^*D(2P_1)$                 &$-6.84$     &\ding{55} &     &$-5.60$  &\ding{55}   \\
					$D_sD_s^*$                    &$-0.01$     &$0.04$    &     &$+0.05$  &$0.10$                  &$D_s^*D_{s1}(2460)$          &$-0.76$     &$0.63$    &     &$-0.54$  &$0.28$   \\
					$D^*D^*$                      &$-0.01$     &$0.09$    &     &$-0.13$  &$0.02$                  &$D^*D_1(2420)$               &$+0.83$     &$2.95$    &     &$+0.23$  &$4.24$ \\
					$D_s^*D_s^*$                  &$+0.02$     &$0.24$    &     &$...$    &$0.14$                  &$D_s^*D_{s1}(2536)$          &$-0.26$     &$0.17$    &     &$-0.05$  &$0.37$   \\
					$D_0(2550)D^*$                &$+1.28$     &$2.20$    &     &$+1.42$  &$4.77$	               &$DD(1^3D_1)$                 &$+0.41$     &$1.76$    &     &$+0.10$  &$0.64$ \\
					$DD_1^*(2600)$                &$+1.01$     &$1.33$    &     &$+1.73$  &$3.76$                  &$DD(1^3D_3)$                 &$+0.57$     &$6.10$    &     &$-0.17$  &$4.95$   \\
\end{tabular}}
		\end{center}
	\end{table*}

	\begin{table}
		\begin{center}
			\caption{The mass shifts and strong decay widths for the $1D$- and $2D$-wave charmonium states (in MeV).
The ``\ding{55}'' is labeled the strong decay channels which are
not open or forbidden. These channels are considered to have no contributions to
the mass shifts and decay widths. The ``$...$'' stands for the predicted
values are negligibly small. }
			\label{1D2D}
			\tabcolsep=0.18cm
			\renewcommand\arraystretch{0.50}
			\scalebox{1.0}{
				\begin{tabular}{ccccccccccccccccc}
					\bottomrule[1.0pt]\bottomrule[1.0pt]
					&\multicolumn{2}{c}{$1^3D_1$}  &  &\multicolumn{2}{c}{$1^3D_3$}
					\\
					\cline{2-3}\cline{5-6}
					&$\Delta M$   &$\Gamma$       &   	 &$\Delta M$   &$\Gamma$           \\
					Channel       &$[3823]$       &$as~\psi(3770)$    &   &$[3847]$   &$as~\psi_3(3842)$   \\
					\bottomrule[0.4pt]
					$DD$               &$-35.12$    &$35.11$          &   &$-4.07$     &$1.24$         \\
					\emph{Total}       &$-35.12$    &$35.11$          &   &$-4.07$     &$1.24$         \\
					$M_{th},\Gamma_{th}$    &\multicolumn{2}{c}{$\bm{3788,35.11}$}          &   &\multicolumn{2}{c}{$\bm{3843,1.24}$}         \\
					$M_{exp},\Gamma_{exp}$  &\multicolumn{2}{c}{$\bm{3774,27.2\pm1.0}$}     &   &\multicolumn{2}{c}{$\bm{3843,2.8\pm0.6}$}    \\
\bottomrule[0.7pt]\bottomrule[0.7pt]
					&\multicolumn{2}{c}{$1^1D_2$}  &  &\multicolumn{2}{c}{$1^3D_2$}
					\\
\end{tabular}}
		\end{center}
	\end{table}

	\begin{table*}
		\begin{center}
			\caption{The mass shifts and strong decay widths for the $3D$-wave charmonium states (in MeV).
The ``\ding{55}'' is labeled the strong decay channels which are
not open or forbidden. These channels are considered to have no contributions to
the mass shifts and decay widths. The ``$...$'' stands for the predicted
values are negligibly small.}
			\label{3D}
			\tabcolsep=0.20cm
			\renewcommand\arraystretch{0.65}
			\scalebox{1.0}{
				\begin{tabular}{cccccccccccccccccccccccc}
					\bottomrule[1.0pt]\bottomrule[1.0pt]
					&&\multicolumn{2}{c}{$3^3D_1$}  &  &\multicolumn{2}{c}{$3^3D_3$}  &   &\multicolumn{2}{c}{$3^1D_2$}  &   &\multicolumn{2}{c}{$3^3D_2$} \\
					 \cline{3-4}\cline{6-7}\cline{9-10}\cline{12-13}
					Channel			&	&$\Delta M$~[4486]  &$\Gamma$~[4463]  &   &$\Delta M$~[4535]  &$\Gamma$~[4461] &
	&$\Delta M$~[4522]  &$\Gamma$~[4461]  &   &$\Delta M$~[4523]  &$\Gamma$~[4468]     \\
					\bottomrule[0.4pt]
					$DD$                       &  &$-0.23$          &$0.86$           &  &$+0.15$      &$0.46$         &   &\ding{55}   &\ding{55}     &   &\ding{55}    &\ding{55} \\
					$D_sD_s$                   &  &$+0.04$          &$0.01$           &  &$-0.11$      &$0.34$         &   &\ding{55}   &\ding{55}     &   &\ding{55}    &\ding{55}  \\
					$DD^*$                     &  &$+0.28$          &$0.01$           &  &$-0.15$      &$2.46$         &   &$-0.81$     &$3.55$        &   &$-0.88$      &$2.96$  \\
					$D_sD_s^*$                 &  &$+0.30$          &$0.94$           &  &$-0.22$      &$0.07$         &   &$-0.12$     &$0.36$        &   &$+0.04$      &$0.53$  \\
					$D^*D^*$                   &  &$-1.64$          &$2.43$           &  &$-0.22$      &$1.20$         &   &$-0.41$     &$1.79$        &   &$-1.18$      &$2.23$  \\
					$D_s^*D_s^*$               &  &$+0.10$          &$3.21$           &  &$-0.38$      &$2.31$         &   &$-0.20$     &$2.02$        &   &$+0.02$      &$1.66$  \\
\end{tabular}}
		\end{center}
	\end{table*}

	\begin{table*}
		\begin{center}
			\caption{The mass shifts and strong decay widths for the $4D$-wave charmonium states (in MeV).
The ``\ding{55}'' is labeled the strong decay channels which are
not open or forbidden. These channels are considered to have no contributions to
the mass shifts and decay widths. The ``$...$'' stands for the predicted
values are negligibly small.}
			\label{4D}
			\tabcolsep=0.20cm
			\renewcommand\arraystretch{0.30}
			\scalebox{1.0}{
				\begin{tabular}{cccccccccccccccccccccccc}
					\bottomrule[1.0pt]\bottomrule[1.0pt]
					&&\multicolumn{2}{c}{$4^3D_1$}  &  &\multicolumn{2}{c}{$4^3D_3$}  &
					&\multicolumn{2}{c}{$4^1D_2$}  &   &\multicolumn{2}{c}{$4^3D_2$} \\
					 \cline{3-4}\cline{6-7}\cline{9-10}\cline{12-13}
					Channel			&	&$\Delta M$~[4767]  &$\Gamma$~[4737]  &   &$\Delta M$~[4823]  &$\Gamma$~[4791] &
	&$\Delta M$~[4808]  &$\Gamma$~[4776]  &   &$\Delta M$~[4808]  &$\Gamma$~[4777]     \\
					\bottomrule[0.4pt]
					$DD$                       &  &$-0.03$       &$0.13$      &  &$+0.04$      &$...$          &   &\ding{55}   &\ding{55}     &   &\ding{55}   &\ding{55}    \\
					$D_sD_s$                   &  &$+0.02$       &$...$       &  &$-0.01$      &$0.07$         &   &\ding{55}   &\ding{55}     &   &\ding{55}   &\ding{55}    \\
					$DD^*$                     &  &$+0.11$       &$...$       &  &$+0.02$      &$0.14$         &   &$-0.16$     &$1.00$        &   &$-0.19$     &$0.87$  \\
					$D_sD_s^*$                 &  &$+0.08$       &$0.19$      &  &$-0.06$      &$0.28$         &   &$-0.08$     &$0.07$        &   &$-0.04$     &$0.06$  \\
					$D^*D^*$                   &  &$-0.39$       &$0.42$      &  &$-0.33$      &$0.61$         &   &$-0.24$     &$0.55$        &   &$-0.35$     &$0.74$  \\
					$D_s^*D_s^*$               &  &$...$         &$0.52$      &  &$+0.03$      &$0.21$         &   &$+0.02$     &$0.25$        &   &$-0.01$     &$0.16$  \\
\end{tabular}}
		\end{center}
	\end{table*}

	\begin{table*}[htbp]
		\begin{center}
			\caption{The mass shifts and strong decay widths for the $5^3D_1$ and $5^3D_3$ charmonium states (in MeV).
The ``\ding{55}'' is labeled the strong decay channels which are
not open or forbidden. These channels are considered to have no contributions to
the mass shifts and decay widths. The ``$...$'' stands for the predicted
values are negligibly small.}
			\label{53D153D3}
			\tabcolsep=0.10cm
			\renewcommand\arraystretch{0.50}
			\scalebox{1.0}{
				\begin{tabular}{cccccccccccccccccc}
					\bottomrule[1.0pt]\bottomrule[1.0pt]
					&\multicolumn{2}{c}{$5^3D_1$}  &  &\multicolumn{2}{c}{$5^3D_3$}&
					&\multicolumn{2}{c}{$5^3D_1$}  &  &\multicolumn{2}{c}{$5^3D_3$}
					\\
\cline{2-3}\cline{5-6}\cline{8-9}\cline{11-12}
					&$\Delta M$~[5029]  &$\Gamma$~[4988]  &   &$\Delta M$~[5091]  &$\Gamma$~[5061] &\emph{Continue}
					&$\Delta M$~[5029]  &$\Gamma$~[4988]  &   &$\Delta M$~[5091]  &$\Gamma$~[5061] \\
					\bottomrule[0.4pt]
					$DD$                              &$+0.01$     &$0.02$   &      &$+0.01$  &$0.02$                   &$DD(1D_2)$                       &$+0.08$     &$1.51$   &      &$+0.16$  &$0.46$ \\
					$D_sD_s$                          &$+0.02$     &$0.01$   &      &$...$    &$0.02$                   &$D_sD_s(1D_2)$                   &$-0.08$     &$0.04$   &      &$-0.10$  &$0.07$ \\
					$DD^*$                            &$+0.04$     &$...$    &      &$+0.01$  &$0.14$                   &$DD(1D'_2)$                      &$-0.72$     &$1.61$   &      &$-0.30$  &$0.35$  \\
					$D_sD_s^*$                        &$+0.03$     &$0.05$   &      &$-0.01$  &$0.02$                   &$D_sD_s(1D'_2)$                  &$-0.18$     &$0.27$   &      &$-0.05$  &$0.04$  \\
					$DD_0(2550)$                      &$+0.20$     &$...$    &      &$+0.09$  &$0.41$                   &$D^*D(1^3D_1)$                   &$-0.99$     &$0.16$   &      &$-0.07$  &$0.53$   \\
					$DD(3^1S_0)$                      &$-1.19$     &$3.00$   &      &$-0.32$  &$0.24$                   &$D_s^*D_s(1^3D_1)$               &$-0.28$     &$0.31$   &      &$-0.03$  &$0.08$  \\
\end{tabular}}
		\end{center}
	\end{table*}

	\begin{table*}[htbp]
		\begin{center}
			\caption{The mass shifts and strong decay widths for the $5^1D_2$ and $5^3D_2$ charmonium states (in MeV).
The ``\ding{55}'' is labeled the strong decay channels which are
not open or forbidden. These channels are considered to have no contributions to
the mass shifts and decay widths. The ``$...$'' stands for the predicted
values are negligibly small.}
			\label{51D253D2}
			\tabcolsep=0.10cm
			\renewcommand\arraystretch{0.50}
			\scalebox{1.0}{
				\begin{tabular}{cccccccccccccccccc}
					\bottomrule[1.0pt]\bottomrule[1.0pt]
					&\multicolumn{2}{c}{$5^1D_2$}  &  &\multicolumn{2}{c}{$5^3D_2$}  &
					&\multicolumn{2}{c}{$5^1D_2$}  &  &\multicolumn{2}{c}{$5^3D_2$}
					\\	 \cline{2-3}\cline{5-6}\cline{8-9}\cline{11-12}
					&$\Delta M$~[5073]  &$\Gamma$~[5044]  &   &$\Delta M$~[5074]  &$\Gamma$~[5030] &\emph{Continue}
					&$\Delta M$~[5073]  &$\Gamma$~[5044]  &   &$\Delta M$~[5074]  &$\Gamma$~[5030] \\
					\bottomrule[0.4pt]
					$DD^*$                            &$-0.04$     &$0.25$    &     &$-0.05$  &$0.20$                      &$DD(1^3D_1)$                     &$+0.08$     &$0.32$    &     &$+0.02$  &$0.14$ \\
					$D_sD_s^*$                        &$-0.02$     &$0.02$    &     &$-0.01$  &$0.01$                      &$D_sD_s(1^3D_1)$                 &$-0.06$     &$0.03$&         &$-0.02$  &$...$  \\
					$D^*D^*$                          &$-0.07$     &$0.15$    &     &$-0.09$  &$0.18$                      &$DD(1^3D_3)$                     &$-0.26$     &$1.15$    &     &$-0.01$  &$1.20$   \\
					$D_s^*D_s^*$                      &$...$       &$0.03$    &     &$-0.01$  &$0.02$                      &$D_sD_s(1^3D_3)$                 &$-0.29$     &$0.20$&         &$-0.19$  &$0.05$   \\
					$D_0(2550)D^*$                    &$-0.29$     &$0.49$    &     &$-0.12$  &$0.27$                      &$DD(1D_2)$                       &$...$       &$...$&          &$+0.01$  &$0.46$ \\
					$D(3^1S_0)D^*$                    &$-0.61$     &$0.09$&         &$-0.63$  &\ding{55}                   &$D_sD_s(1D_2)$                   &$...$       &$...$&          &$-0.06$  &$0.02$ \\
\end{tabular}}
		\end{center}
	\end{table*}

\section*{Acknowledgements }
	
We thank useful discussions from Long-Cheng Gui, Qi-Fang L\"{u}, Xiang Liu, Zhi-Yong Zhou, Ying Chen, and Qiang Zhao.
This work is supported by the National Natural Science Foundation of China under Grants Nos. 12235018, 12175065, 12205216.

\bibliographystyle{unsrt}

\end{document}